\documentstyle[rotating,epsfig,graphicx]{mn2e}

\def\vsini{$V\!\sin i$}
\def\teff{T$_{\rm{eff}}$}

\def\logg{$\log~g$}

\def\kps{km~s$^{-1}$}

\def\logl{$\log~L/L_{\odot}$}
\def\lage{$\log~{\rm Age}$}
\def\mmsun{$M/M_{\odot}$}

\title[Spectroscopic analysis of southern B and Be stars]{Spectroscopic analysis of southern B and Be stars}
\author[R. S. Levenhagen \& N. V. Leister]{R. S. Levenhagen$^{1}$\thanks{E-mail:
savarino@astro.iag.usp.br (RSL)} and N. V. Leister$^{1}$\footnotemark[1]\thanks{This study is based on observations made at ESO La Silla, Chile
and MCT/LNA, Brazil.}\\
$^{1}$Instituto de Astronomia, Geof\'{\i}sica e Ci\^ encias Atmosf\'ericas da Universidade de S\~ao Paulo,
CUASO 05508-900 S\~ao Paulo, Brazil\\}

\begin{document}

\date{Accepted 1988 December 15. Received 1988 December 14; in original form 1988 October 11}

\pagerange{\pageref{firstpage}--\pageref{lastpage}} \pubyear{2005}

\maketitle

\label{firstpage}

\begin{abstract}
Spectroscopic monitoring of 141 southern field B type stars, 114 of them known to 
exhibit the Be phenomenon, allowed the estimation of their projected rotational 
velocities, effective temperatures and superficial gravities from both line and 
equivalent width fitting procedures. Stellar ages, masses and bolometric luminosities 
were derived from internal structure models. Without taking into account for the 
effects of gravity darkening, we notice the occurrence of the Be phenomenon in 
later stages of main sequence phase. 
\end{abstract}


\section{Introduction}

\indent Despite the large amount of works on the subject of B and Be stars 
published in the past years, both theoretical and observational, there are 
still many remaining unsolved problems concerning their outstanding 
peculiarities. Among them the origin of their high rotation velocities, 
dependence of Be frequency counts with evolutionary 
stages in the main sequence, presence and origin of magnetic fields and mass 
loss (Vinicius et al. 2006, McSwain \& Gies 2005, 
Levenhagen \& Leister 2004, Levenhagen et al 2003, Porter \& Rivinius 2003).

\indent Concerning particularly the case of Be stars, the most outstanding observable 
characteristic in the visible domain is the existence of emission in Balmer 
lines, being often accompanied by emission of singly ionized metals and changes 
of the spectrophotometric characteristics (Moujtahid et al. 1999; Levenhagen 2004). In these 
stars, emissions and energy distributions are variable and the most important 
among these variations are ``phase transitions'', which most likely reflect 
changes in the structure of the circumstellar envelope (CE) and processes 
related with its formation.

\indent One of the fundamental questions related to the Be phenomenon concerns the origin 
of the fast rotation rates of the central star. Rotation has quite different effects 
on stellar evolution, as changes of internal hydrostatic equilibrium, transport of 
chemical species by meridional circulation and effects on mass loss rates (Maeder 1999).

These days it is a consense that fast surface rotation plays a crucial role in triggering 
the Be phenomenon, where the onset of new observational evidences indicate that these objects are in fact near critical rotators, with $\omega \simeq 0.88$ (Zorec, Fr\'emat \& Cidale 2005, 
Fr\'emat et al. 2005, Townsend, Owocki \& Howarth 2004).

There are actually three main possibilities for that:
(i) it would be a permanent innate property; (ii) it would be due to spin up by binary mass 
transfer; (iii) or it would be acquired somehow during the main sequence evolution phase 
(Crampin \& Hoyle 1960, Schild \& Romanishin 1976). Early attempts to describe the Be 
phenomenon suggested the occurrence of Be stars during the secondary contraction phase 
(Schmidt-Kaler 1964). A few years later, it was seen the appearance of a significant fraction 
of Be stars close to the ZAMS (Schild \& Romanishin 1976). In the 80's it was generally 
accepted that those objects occupy the whole main sequence band at different evolutionary 
phases (Mermilliod 1982; Slettebak 1985).

 \indent Maeder, Grebel \& Mermilliod (1999) explained the origin of high rotation rates by 
two possible scenarios of low metallicity effects in star formation regions. In the first 
scenario a weak coupling of magnetic fields with star forming matter leads to less angular 
momentum losses and a consequent spin-up. The second possibility is based on a low opacity 
effect in the pre main sequence phase which in turn determines the external convective zones 
to fade away and thus leading to less angular momentum losses. 

 Today some works present the Be phenomenon as a result of an evolutionary effect instead 
(Fabregat \& Torrej\'on 2000). In a recent work on Be stars in open clusters, McSwain 
\& Gies (2005) pointed out that the Be phenomenon is not strongly dependent on metallicity 
or cluster density, instead it is influenced by the evolutionary stage.
One problem with those studies is that they rely on photometric measurements only, which tends, 
in the case of Be stars, to systematically overestimate the magnitudes due to the presence 
of a circumstellar environment. On the other hand, Zorec et al (2006) determined photospheric
parameters of a sample of 97 field Be stars near the Sun based on spectroscopic measurements 
considering both rotating and non-rotating model atmosphere scenarios. In this work they found 
that, without taking into account for the effects of fast rotation in the models, Be stars usually
appear near the TAMS rather than the ZAMS in the HR diagram, i.e. the Be phenomenon would arise 
due to an apparent evolutionary effect. Later on, considering the effects of gravity darkening,
the stellar distribution in the HR diagram becomes more homogeneous, and Be stars are allowed
to occur at any stages within the main sequence phase. In order to elucidate this problem 
further investigations are still needed.

\indent In this paper we present the results of spectroscopic analyses carried out with 141 
southern B stars (where 114 are field Be stars), some of them located away from the solar 
neighbourhood, in order to provide physical parameters estimations of \vsini, \teff~ and 
\logg, and interpolations in the tabulated evolutionary tracks by Schaller et al. (1992) 
necessary for further investigations on these objects. 
These parameters have been estimated from NLTE model atmospheres without taking into account 
for gravitational darkening (von Zeipel 1924a,b) and their positions in the HR diagram were 
established. We also estimated the errors committed in the physical parameters (\teff,\logg) 
by neglecting the rotational effects.

\section{Observations}

 \indent High resolution and signal-to-noise spectroscopic observations were carried out in 
the Southern Hemisphere both with FEROS spectrograph associated to the 1.52m telescope at 
ESO/La Silla (Chile) and with Coud\'{e} spectrograph at the 1.60m telescope of MCT/LNA 
(Brazil). We made observations of B and Be stars over 5 years, from October/1999 to July/2003. 
ESO spectra were taken with a spectral coverage of 3560-9200 \AA, with typical S/N $\sim 200$
and a reciprocal dispersion of 0.09~ \AA/pixel, concerning the He\,{\sc i}4471~\AA~ line. 
LNA spectra were gathered with two CCD cameras: a WI101 CCD from 4279~\AA~ to 4720~\AA~ with 
reciprocal dispersion of 0.43~ \AA/pixel (at 4471~\AA~) and a WI098 CCD from 3939~\AA~ to 5060~\AA~ with a 
reciprocal dispersion of 0.24~ \AA/pixel (at 4471~\AA~) using a 600 groove $\rm mm^{-1}$ grating blazed at
4500~\AA~ in first order and typical S/N ratio $\sim 150$ (For more details see Table 1). The images were 
processed in the same sense, basically starting with bias and overscan subtraction, followed by the division 
of the stellar spectra by an average flat-field exposure, compiled from all the flat fields taken during 
the night. The spectra were then wavelength-calibrated and corrected for telluric line contamination. 
Standard stars of rotation velocity and flux were also observed. All the mentioned data reduction was 
accomplished using the IRAF  \footnote{IRAF is distributed by the National Optical Astronomy Observatories, 
which is operated by the Association of Universities for Research in Astronomy (AURA), Inc., under cooperative 
agreement with the National Science Foundation} package.

\addtocounter{table}{+0}%
\begin{table*}
\caption{Spectroscopic observations of target objects.}
\centering
\tiny{
\begin{tabular}{l|ccccllc}
\hline\hline
Object  &  $\ell$ (o)   &  $\rm b$ (o)  &  $\alpha$ (J2000)  & 
$\delta$ (J2000)  &  Telescope & Epoch  &   Wavelength range  (\AA) \\
\hline\hline
HD 10144 & 290.84 & -58.79 &	1	37	42	&	-57	14	12 & ESO  1.52 m	& October/1999  & 3560 - 9200 \\
HD 14850 & 225.38 & -69.74 &	2	23	0	 	& 	-29	37	10	& ESO 1.52 m     & October/2001  &	3560 - 9200 \\
HD 15371 & 267.12 & -62.24 &	2	26	59	& 	-47	42	13	& LNA 1.60 m & July/2003    & 3939 - 5060 \\
HD 16582  & 170.76 & -52.21 &  2   39	28 & 0	      19	42 & LNA 1.60 m & July/2003    & 3939 - 5060	\\
HD 17891  & 143.59 & -10.73 & 02   54    00 & 47      09    39 & ESO 1.52 m     & October/2001  & 3560 - 9200 \\
HD 20340  & 203.36 & -55.13 & 03   15    45 & -16     49    43 & ESO 1.52 m     & October/2001  & 3560 - 9200 \\
HD 28248  & 267.15 & -42.23 & 4	24	6  & -57	15	11 & ESO 1.52 m     & September/2002 & 3560 - 9200\\
HD 29557  & 223.65 & -39.36 & 4    38    16 & -24     39    30 & ESO 1.52 m     & October/2001  & 3560 - 9200 \\
HD 33453  & 245.40  & -36.21 & 5	8	26 & -40	54	44 & ESO 1.52 m     & October/2001  & 3560 - 9200\\
HD 33599  & 271.32 & -35.91 & 5	7	12  & -61	48	18  & ESO 1.52 m     & October/2001  & 3560 - 9200\\
HD 35165  & 238.07 & -32.61 & 5	21	16 & -34	20	42 & ESO 1.52 m     & October/2001  & 3560 - 9200\\
HD 35468  & 196.93 & -15.95 & 5	25	7  & 6	      20	58 & ESO 1.52 m     & September/2002 & 3560 - 9200\\
HD 36012  & 201.19 & -17.25 & 5	28	48 & 2	       9	52 & ESO 1.52 m     & October/2001  & 3560 - 9200\\
HD 36861  & 195.05 & -11.99 & 5	35	8 	  & 9  	56	5 	 & ESO 1.52 m     & September/2002 & 3560 - 9200\\
HD 37023  & 209.01 & -19.38 & 5	35	17 & -5	23	15  & ESO 1.52 m     & September/2002 & 3560 - 9200\\
HD 37490  & 200.73 & -14.03 & 5    39    11 &  4      07    17& ESO 1.52 m     & April/2001    & 3560 - 9200\\
HD 37795  & 238.81 & -28.86 & 5	39	38 & -34	4	26& ESO 1.52 m     & October/2001  & 3560 - 9200\\
HD 37935  & 276.51 & -32.10  & 5	36	55      & -66	33	37     & ESO 1.52 m     & October/2001  & 3560 - 9200\\
HD 43122  & 244.10  & -23.20  & 6	12	49  & -37	4	53& ESO 1.52 m     & October/2001  & 3560 - 9200\\
HD 43285  & 203.42 & -5.13  & 6	15	40 & 6	      3	58& ESO 1.52 m     & October/2001  & 3560 - 9200\\
HD 43544& 224.13& -15.15& 06 16 07& -16 37 04& ESO 1.52 m     & October/2001 & 3560 - 9200\\
HD 43789& 252.20& -24.71& 06 15 56& -44 37 10& ESO 1.52 m     & October/2001 & 3560 - 9200\\
HD 44743& 226.06& -14.27& 06 22 41& -17 57 21& ESO 1.52 m     & October/2001 & 3560 - 9200\\
HD 44996& 221.58& -11.80& 06 24 20& -12 57 42& ESO 1.52 m     & October/2001 & 3560 - 9200\\
HD 45871& 240.47& -18.59& 06 28 39& -32 22 16& ESO 1.52 m     & October/2001 & 3560 - 9200\\
HD 46131& 230.96& -14.35& 06 30 38& -22 19 18& ESO 1.52 m     & October/2001 & 3560 - 9200\\
HD 46380& 217.53&  -7.56& 06 32 43& -07 30 32& ESO 1.52 m     & October/2001 & 3560 - 9200\\
HD 47839& 202.94&  +2.20& 06 40 58& +09 53 44& ESO 1.52 m     & October/2001 & 3560 - 9200\\
HD 48099& 206.21&  +0.80& 06 41 59& +06 20 43& ESO 1.52 m     & October/2001 & 3560 - 9200\\
HD 48282& 221.28&  -6.79& 06 42 12& -10 29 53& ESO 1.52 m     & October/2001 & 3560 - 9200\\
HD 49131& 240.50& -14.73& 06 45 31& -30 56 56& ESO 1.52 m     & October/2001 & 3560 - 9200\\
HD 49319& 248.84& -17.84& 06 46 03& -39 32 24& ESO 1.52 m     & October/2001 & 3560 - 9200\\
HD 49330& 211.84&  -0.42& 06 47 57& +00 46 34& ESO 1.52 m     & October/2001 & 3560 - 9200\\
HD 49336& 247.13& -17.18& 06 46 12& -37 46 31& ESO 1.52 m     & October/2001 & 3560 - 9200\\
HD 50013& 242.36& -14.49& 06 49 50& -32 30 30& ESO 1.52 m     & April/2000   & 3560 - 9200\\
HD 50209& 213.28&  +0.03& 06 52 10& -00 17 43& ESO 1.52 m     & October/2001 & 3560 - 9200\\
HD 50696& 213.10&  +0.74& 06 54 22&+00 10 54 & ESO 1.52 m     & October/2001 & 3560 - 9200\\
HD 50737& 224.98&  -5.44& 06 53 52& -13 11 09& ESO 1.52 m     & October/2001 & 3560 - 9200\\
HD 50850& 229.61&  -7.65& 06 54 09&-18 17 10 & ESO 1.52 m     & October/2001 & 3560 - 9200\\
HD 51309& 228.70&  -6.68& 06 56 08& -17 03 15& ESO 1.52 m     & October/2001 & 3560 - 9200\\
HD 52159& 223.81&  -3.25& 06 59 42& -11 09 26& ESO 1.52 m     & October/2001 & 3560 - 9200\\
HD 52244& 228.33&  -5.52& 06 59 46& -16 12 02& ESO 1.52 m     & October/2001 & 3560 - 9200\\
HD 55606& 217.31&  +3.97& 07 13 34&-02 04 39 & ESO 1.52 m     & October/2001 & 3560 - 9200\\
HD 58715& 209.52& +11.68& 07 27 09& +08 17 21& LNA 1.60 m & May/2001    & 4279 - 4720 \\
HD 59868& 257.26& -12.53& 07 29 48& -44 54 42& ESO 1.52 m     & October/2001 & 3560 - 9200\\
HD 63150& 251.17&  -5.90& 07 45 55&-36 29 53 & LNA 1.60 m & May/2001    & 4279 - 4720 \\
HD 67698& 242.66&  +4.92& 08 08 19& -23 37 04& ESO 1.52 m     & April/2001   & 3560 - 9200\\
HD 70461& 246.56&  +5.87& 08 21 13&-26 19 59 & ESO 1.52 m     & April/2001   & 3560 - 9200\\
HD 74280& 223.25& +26.32& 08 43 13& +03 23 55& LNA 1.60 m & May/2001    & 4279 - 4720 \\
HD 79447& 280.22&  -9.61& 09 11 16& -62 19 01& LNA 1.60 m & May/2001    & 4279 - 4720 \\
HD 90177& 285.15&  -1.98& 10 22 53& -59 37 28& LNA 1.60 m & May/2003    & 3939 - 5060\\
HD 100546 & 296.37 & -8.32 &	11	33	25	&	-70	11	41 & LNA 1.60 m & June/2003    & 3939 - 5060 \\
HD 100889 & 274.25 & 48.86 &	11	36	40	&	-9	48	8	 & LNA 1.60 m & May/2001     & 3939 - 5060 \\
HD 104582 & 296.73 & 2.95 &	12	2	35	&	-59	20	13	 & ESO  1.52 m    & August/2001   &	3560 - 9200 \\
HD 105435 & 296.00 & 11.57 & 	12	8	21	&	-50	43	20 & LNA 1.60 m & May/2003     & 3939 - 5060 \\
HD 105521 & 294.39 & 20.93 &	12	8	54	&	-41	13	53 & LNA 1.60 m & July/2003    & 3939 - 5060	\\
HD 105937 & 296.78 & 10.03 &	12	11	39	&	-52	22	6	 & LNA 1.60 m	& February/2003& 3939 - 5060	\\
HD 106309 & 298.18 & 3.13  & 12    14    01     &     -59   23    48 & LNA 1.60 m & May/2001     & 4279 - 4720 \\    
HD 106793 & 298.15 & 6.14  & 12    16    59      &     -56   24    24  & LNA 1.60 m & May/2001     & 4279 - 4720 \\   
HD 110432 & 301.96 & -0.20 & 	12	42	50	&	-63	3	31 & LNA 1.60 m & July/2003    & 3939 - 5060	\\
HD 110699 & 302.05 & 3.72 & 	12	44	37	&	-59	9	36	 & ESO 1.52 m   	& August/2001   & 3560 - 9200	\\
HD 112078 & 303.35 & 3.72 & 	12	54	39	&	-59	8	48 & LNA 1.60 m & July/2003    & 3939 - 5060	\\
HD 112091 & 303.37 & 5.70 & 	12	54	36	&	-57	10	7	 & LNA 1.60 m & July/2003    &	3939 - 5060	\\
HD 112107 & 303.44 & 10.38 &	12	54	41	&	-52	29	2	 & ESO  1.52 m    & April/2001    &	3560 - 9200 \\
HD 112512 & 303.78 & 3.78  & 12    58    00     &     -59   05    03 & LNA 1.60 m & May/2001     & 4279 - 4720 \\
HD 113120 & 303.87 & -8.63 &	13	3	5	&	-71	28	32 & LNA 1.60 m & July/2003    & 3939 - 5060	\\
HD 118094 & 307.98 & -0.71 &	13	36	20	&	-63	8	44 & LNA 1.60 m & July/2003    & 3939 - 5060	\\
HD 119423 & 308.22 & -4.44 &	13	45	18	&	-66	45	16 & ESO  1.52 m  	& July/2002    &	3560 - 9200 \\
HD 119682 & 309.16 & -0.72 &	13	46	32	 	&	-62	55	24.13	 & ESO  1.52 m    & July/2002    &	3560 - 9200 \\
HD 120324 & 314.24 & 19.12 &	13	49	36	&	-42	28	25 & LNA 1.60 m & July/2003    & 3939 - 5060	\\
\hline
\end{tabular}}
\end{table*}

\eject

\addtocounter{table}{-1}%
\begin{table*}
\caption{Continued.}
\centering
\tiny{
\begin{tabular}{l|ccccllc}
\hline\hline
Object  &  $\ell$ (o)   &  $\rm b$ (o)  &  $\alpha$ (J2000)  & 
$\delta$ (J2000)  &  Telescope & Epoch  &   Wavelength range  (\AA) \\
\hline\hline
HD 120991 & 313.84 & 14.42 &	13	53	57	&	-47	7	41 & LNA 1.60 m & May/2001     & 4279 - 4720 \\
HD 124639 & 305.94 & -20.53 & 14	24	22	&	-82	50	53	 & ESO 1.52 m     & July/2002    & 3560 - 9200	\\
HD 126527 & 316.38 & 4.86 & 	14	27	56	&	-55	28	10	 & ESO 1.52 m   	& August/2001   & 3560 - 9200	\\
HD 126986 & 317.93 & 7.71 & 	14	30	32	&	-52	15	20	 & ESO 1.52 m     & August/2001   & 3560 - 9200	\\
HD 127112 & 315.59 & 1.49 & 	14	31	45	&	-58	53	22 & ESO 1.52 m     & August/2001   & 3560 - 9200 \\
HD 127208 & 331.15 & 34.91 &	14	30	40	&	-22	27	39 & ESO 1.52 m     & April/2001    &	3560 - 9200 \\
HD 127381 & 318.93 & 9.25 & 	14	32	37	&	-50	27	25 & LNA 1.60 m & May/2001     & 4279 - 4720 \\
HD 127972 & 322.77 & 16.67 &	14	35	30	&	-42	9	28 & ESO 1.52 m     & April/2000    & 3560 - 9200 \\
HD 129956 & 353.72 & 51.97 &	14	45	30	&	0	43	2	 & LNA 1.60 m & May/2001     & 4279 - 4720 \\
HD 130437 & 317.22 & -0.77 &	14	50	50	&	-60	17	10 & ESO 1.52 m     & April/2000    & 3560 - 9200 \\
HD 130534 & 320.76 & 6.43 &	14	50	41	&	-52	15	52 & ESO 1.52 m     & August/2001   & 3560 - 9200	\\
HD 131168 & 324.10 & 11.94 & 	14	53	22	&	-45	51	20	 & ESO 1.52 m     & April/2001    & 3560 - 9200 \\
HD 132947 & 317.37 & -4.06 &	15	4	56	&	-63	7	52	 & ESO 1.52 m     & April/2000    & 3560 - 9200 \\
HD 134401 & 316.69 & -6.96 &	15	13	12	&	-65	58	9	 & ESO 1.52 m     & August/2001   & 3560 - 9200 \\
HD 134481 & 325.46 & 7.89 & 	15	11	56	&	-48	44	16 & LNA 1.60 m & July/2003    & 3939 - 5060	\\
HD 134671 & 324.57 & 6.09 &	15	13	2  	&	-50	44	6	 	& ESO 1.52 m     & April/2001    &	3560 - 9200 \\
HD 135734 & 326.86 & 8.05 &	15	18	32	&	-47	52	30	& LNA 1.60 m & July/2003    &	3939 - 5060	\\
HD 136968 & 324.60 & 2.47 &  15    25    55      &     -53   46    15       & LNA 1.60 m & May/2001     &  4279 - 4720 \\
HD 137387 & 313.85 & -14.01 &15	31	30	&	-73	23	22	& LNA 1.60 m & July/2003    &	3939 - 5060	\\
HD 137518 & 329.81 & 9.40 &	15	28	17	& 	-45	8	5	 	& ESO 1.52 m     & April/2001    &	3560 - 9200 \\
HD 142237 & 327.41 & -1.11 &	15	56	5	 	& 	-54	57	9	 	& ESO 1.52 m     & April/2001    &	3560 - 9200 \\
HD 142349 & 325.82 & -3.21 &	15	57	7	 	& 	-57	34	48	 	& ESO 1.52 m     & August/2001   &	3560 - 9200 \\
HD 143545 & 332.29 & 2.91 &	16	3	9	 	& 	-48	43	21	 	& ESO 1.52 m     & August/2001   &	3560 - 9200 \\
HD 143578 & 341.31 & 13.00 & 16    2     38      &     -35   15    11       & ESO 1.52 m     & April/2001    &    3560 - 9200 \\ 
HD 143700 & 333.23 & 3.74 &	16	4	2	& 	-47	28	33	& ESO 1.52 m     & August/2001   &	3560 - 9200 \\
HD 144555 & 330.81 & -0.12 &	16	8	55	& 	-51	57	43	 	& ESO 1.52 m     & August/2001   &	3560 - 9200 \\
HD 146531 & 327.66 & -5.69 &	16	19	55	& 	-58	10	1	 	& ESO 1.52 m     & August/2001   &	3560 - 9200 \\
HD 149757 & 6.28 & 23.59 &	16	37	9	& 	-10	34	1	 	& ESO 1.52 m     & April/2000    &	3560 - 9200 \\
HD 150193 & 355.60 & 14.83 &	16	40	17	& 	-23	53	45	 	& ESO 1.52 m     & April/2000    &	3560 - 9200 \\
HD 150288 & 338.13 & -0.54 &	16	42	16	& 	-47	0	56	 	& ESO 1.52 m     & August/2001   &	3560 - 9200 \\
HD 150422 & 338.61 & -0.31 &	16	43	7	 	& 	-46	30	23	 	& ESO 1.52 m     & August/2001   &	3560 - 9200 \\
HD 151113 & 339.90 & -0.21 &	16	47	30	& 	-45	27	37	& ESO 1.52 m     & August/2001   &	3560 - 9200 \\
HD 152060 & 343.65 & 1.62 &	16	52	59	& 	-41	24	27	 	& ESO 1.52 m     & September/2001 & 3560 - 9200\\
HD 152478 & 336.78 & -4.64 &	16	56	8	& 	-50	40	29	& ESO 1.52 m     & August/2001   & 3560 - 9200\\
HD 152979 & 340.64 & -2.18 &	16	58	56	& 	-46	7	44	& ESO 1.52 m     & August/2001   & 3560 - 9200\\
HD 153199 & 343.04 & -0.52 &	16	59	56	& 	-43	13	14	 	& ESO 1.52 m     & April/2001    & 3560 - 9200 \\
HD 154154 & 339.57 & -4.54 &	17	6	7	 	& 	-48	25	7	 	& ESO 1.52 m     & August/2001   & 3560 - 9200\\
HD 155851 & 353.32 & 3.33 &	17	15	33	& 	-32	41	23	 	& ESO 1.52 m     & April/2001    & 3560 - 9200\\
HD 156325 & 353.77 & 2.93 &	17	18	20	& 	-32	33	11	& ESO 1.52 m     & April/2001    & 3560 - 9200\\
HD 156702 & 349.05 & -0.98 &	17	20	50	& 	-38	39	8	 	& ESO 1.52 m     & July/2002    & 3560 - 9200\\
HD 158427 & 340.76 & -8.83 & 17    31    50     &     -49   52    34      & ESO 1.52 m     & April/2000    & 3560 - 9200 \\  
HD 159489 & 345.28 & -7.10 &	17	37	13	& 	-45	9	26	 	& ESO 1.52 m     & August/2001   & 3560 - 9200\\
HD 160202 & 356.59 & -0.69 &	17	40	1	& 	-32	12	3	 	& ESO 1.52 m     & April/2001    & 3560 - 9200\\
HD 161774 & 356.14 & -3.14 & 17    48    51     &     -33   51    45      & LNA 1.60 m & May/2001     & 4279 - 4720 \\    
HD 164284 & 30.99 & 13.37 &	18	0	15	& 	4	22	7	 	& LNA 1.60 m & July/2003    & 3939 - 5060	\\
HD 164816 & 6.06 & -1.20 &	18	3	56 & 	-24	18	45	 	& LNA 1.60 m & June/2003    & 3939 - 5060	\\
HD 164906 & 6.05 & -1.33 &	18	4	25	& 	-24	23	8	 	& ESO 1.52 m     & July/2002    & 3560 - 9200\\
HD 164947 & 6.11 & -1.36 &	18	4	41	 	& 	-24	20	54	 	& LNA 1.60 m & June/2003    & 3939 - 5060	\\
HD 165052 & 6.12 & -1.48 &	18	5	10	& 	-24	23	54	& ESO 1.52 m     & October/2001  & 3560 - 9200\\
HD 166566 & 14.51  & 1.38   & 18	11	50 & -15	40	47 & ESO 1.52 m     & September/2002 &  3560 - 9200\\
HD 170235 & 7.95   & -6.73  & 18	29	21 & -25	15	23 & ESO 1.52 m     & August/2001   &	3560 - 9200 \\
HD 170835 & 13.67  & -4.58  & 18	32	16  & -19	13	1   & ESO 1.52 m     & August/2001   &  3560 - 9200\\
HD 171054 & 18.49  & -2.34  & 18	33	8	  & -13	54	43  & ESO 1.52 m     & September/2002 & 3560 - 9200\\
HD 171219 & 35.73  & 6.51   & 18	33	17  & 5	      26	43 & ESO 1.52 m     & July/2002    & 3560 - 9200\\
HD 172256 & 11.40   & -7.78  & 18	40	11 & -22	39	49 & ESO 1.52 m     & August/2001   & 3560 - 9200\\
HD 173948 & 333.61 & -23.87 & 18   52    13 & -62     11    15 & LNA 1.60 m & July/2003    & 3939 - 5060	\\
HD 174513 & 25.95  & -3.49  & 18	51	9  & -7	47	55 & ESO 1.52 m     & July/2002    & 3560 - 9200\\
HD 174705 & 22.64  & -5.46  & 18   52    16  & -11     37    57  & LNA 1.60 m & May/2001     & 4279 - 4720 \\
HD 179253 & 355.25 & -21.80  & 19	13	44  & -42	24	48  & ESO 1.52 m     & April/2001    & 3560 - 9200\\
HD 183914 & 62.12  & 4.57   & 19	30	45 & 27	57	54 & LNA 1.60 m & July/2003    & 3939 - 5060	\\
HD 184279 & 41.15  & -7.62  & 19	33	36 & 3	      45	40 & ESO 1.52 m     & July/2002    & 3560 - 9200\\
HD 185037 & 70.58  & 7.89   & 19	35	48 & 36	56	40 & LNA 1.60 m & July/2003    & 3939 - 5060	\\
HD 198001 & 37.68  & -30.10  & 20	47	40 & -9	29	44 & LNA 1.60 m & May/2001     & 4279 - 4720 \\
HD 205637 & 31.94  & -44.99 & 21   37    04 & -19     27    57 & LNA 1.60 m & July/2003    & 3939 - 5060	\\
HD 208886 & 15.64  & -52.59 & 21	59	55 & -31	31	32  & LNA 1.60 m & May/2001     & 4279 - 4720 \\
HD 217891 & 78.79  & -49.61 & 23	3	52  & 3	      49	12 & LNA 1.60 m & July/2003    & 3939 - 5060	\\
HD 224686 & 311.30  & -50.71 & 23	59	54 & -65	34	37 & LNA 1.60 m & July/2003    & 3939 - 5060	\\
HD 316341 & 359.46 & -1.08  & 17	48	35  & -29	57	28  & ESO 1.52 m     & April/2001    & 3560 - 9200\\
HD 316587 & 359.89 & -1.75  & 17	52	13  & -29	55	52  & ESO 1.52 m     & April/2001    & 3560 - 9200\\
HD 322422 & 344.99 & 1.63   & 16	57	23  & -40	21	39   & ESO 1.52 m     & August/2001   & 3560 - 9200\\
HD 330950 & 335.40  & -1.30   & 16	34	43  & -49	33	9   & ESO 1.52 m     & September/2002 & 3560 - 9200\\
\hline
\end{tabular}}
\end{table*}

\subsection{Spectroscopic analysis and results}

In this work we derive the (\teff,\logg) parameters with the use of non-LTE model 
atmospheres neglecting the effects of gravity darkening on the shape of line profiles. 
As Be stars are fast rotators, these photospheric quantities actually reflect only the 
average apparent photospheric characteristics, ie. those of the observed stellar 
hemisphere deformed by the fast rotation (Fr\'emat et al. 2005). 

 The determination of stellar temperatures, gravities and rotation velocities is 
achieved in this work in basically three steps: 

(i) First guess of \vsini~ from the average of Fourier Transforms (FT) of all neutral 
helium lines in the domain 4000 to 5000 ~\AA. To this end we adopted a quadratic 
limb-darkening law (Wade \& Rucinski 1985) to determine the rotational broadening 
function (Gray 1992; Carroll 1933) of the form:

\begin{equation}
G(y)=C_{1}(1-y^{2})^{1/2} + C_{2}(1-y^{2}) + C_{3}(1-y^{2})^{2}  
\end{equation}

where $C_{1}$, $C_{2}$ and $C_{3}$ are constants dependent on the linear and quadratic 
limb-darkening coefficients. The Fourier Transform of the broadening function can be 
written as:

\begin{eqnarray}
g(\alpha) &=& 2\int_{0}^{1}{G(y)\cos{\sigma b y}dy} = \nonumber \\
&=& C_{1} \pi {{J_{1}(\alpha)}\over{\alpha}} \left( 1 + {{6}\over{\alpha^2}} {{C_{3}}\over{C_{1}}} \right) + \nonumber \\ 
&+& C_{2} \left( {{2}\over{\alpha}} \right) ^{2} \left( {{\sin{\alpha}}\over{\alpha}} -
\cos{\alpha} \right) - 3\pi C_{3} {{J_{0}(\alpha)}\over{\alpha^{2}}}
\end{eqnarray}

The rotation velocity values are then estimated from the ratio of the first zero of the 
broadening function's TF ($\alpha_{1}$) and the first zero of the TF of the observed 
profile ($\nu_{1}$) as:

\begin{equation}
V\!\sin i = {{c}\over{\lambda_{0}}}{{\alpha_{1}}\over{2\pi\nu_{1}}}
\end{equation}

This method is suitable only for a first guess on \vsini~ values (Levenhagen 2004, Levenhagen 
\& Leister 2004), but takes into account only for limb-darkening coefficients related to the continuum spectrum and the effect of gravity darkening is not considered; 

(ii) First estimation of effective temperature and superficial gravity through the ionization 
equilibria of He\,{\sc ii}/He\,{\sc i}, Si\,{\sc iii}/Si\,{\sc ii}, O\,{\sc iii}/O\,{\sc ii}, 
N\,{\sc iii}/N\,{\sc ii} and Balmer equivalent width fitting, constraining the initial solutions 
to some determined region in the (\teff, \logg) plane. For this sake it were used LTE atmosphere 
models from Kurucz (1992). This process originates a series of curves in the (\teff, \logg) plane 
which intercept each other delimiting regions of possible solutions where the best one is to be 
given by its baricentre. In order to choose which region is well suitable to a given object
we compare each baricentre value with (\teff, \logg) values determined from photometric Johnson's 
UBV, Str\"omgren uvby and H$\beta$ values for that specific object in the literature, and taking 
into account its spectral classification. In the particular case of Be stars, the presence of 
circumstellar environment (CE) emission causes an overestimation of $m_{v}$ magnitudes. An attempt 
to clear up this effect was to employ the corrections proposed by Zorec \& Briot (1997) for each 
subspectral type; 

(iii) Once established a first guess of (\teff, \logg, \vsini) parameters, we continue with a more 
detailed line fitting procedure with theoretical profiles synthesized with the SYNSPEC fortran 
program (Hubeny, Hummer, \& Lanz 1994) in the spectral range 4000 to 5000~\AA~ from a grid of 
non-LTE stellar atmosphere models generated with TLUSTY code (Hubeny 1988). The fitting procedure 
is performed with models calculated for a parameter space around the starting (\teff,\logg,\vsini) 
set using the downhill simplex algorithm (Nelder \& Mead 1965). Figure 1 shows the fittings of 
He\,{\sc i} 4471 \AA~ and Mg\,{\sc ii} 4481 \AA~ line profiles for 8 sample stars. The same fitting
procedure was adopted for all 141 B/Be objects. In Figure 2 we show more detailed fittings achieved 
to the observed spectrum of Achernar (HD 10144) in wavelength regions near the Balmer (H$\beta$, 
H$\gamma$, H$\delta$) lines. In some observed slow rotating Be stars where $V\!\sin i \la 
100$ \kps, which are probably seen pole-on, the emission in the H$\gamma$ line profile is often 
present. This emission makes uncertain the line fitting procedure with models. In such cases we used 
the method outlined in Chauville et al. (2001). Choosing two points $x_1$ and $x_2$ in the spectrum 
(with $x_{i} = \lambda_{i} - \lambda{c}$), a third point is given by:

\begin{equation}
x_3 = \left( x_1 x_2 \right) ^{1/2}
\end{equation}

\noindent By definition the ordinate values related to $x_1$, $x_2$ and $x_3$ are given through:

\begin{equation}
y_i = - \left[ \ln{\psi^{\rm obs} \left( \lambda_{i} \right)} \right] ^{-1}
\end{equation}

\noindent where $\psi^{\rm obs}$ is the observed profile. The final adjusted profile
$\psi(\lambda)$ is of the form:

\begin{equation}
\psi(\lambda) = \exp{ \left\lbrace - \left[ a\; \left( \lambda - \lambda_c \right) ^{b} + c \right] ^{-1} \right\rbrace  }
\end{equation}

\noindent where the constants $a$, $b$ and $c$ are related by:

\begin{equation}
\left\{\begin{array}{l} c = {{(y_{1}y_{2}-y_{3}^{2})}\over{(y_{1}+ y_{2}-2y_{3})}} \\
\ln{ \left( y_i - c \right) } = b\;\ln{ x_i } + a  \\ 
\end{array}\right.
\end{equation}

Relation (6) is used to determine the absorption-like underlying photospheric 
component of the Balmer line seen in emission, or in the worst cases where the 
star is in full emission we estimated the \teff~ and \logg~ parameters using 
only the H$\delta$ profile.\par

To perform the operation outlined in step (ii) we use, whenever possible, lines in two neighbouring 
ionization states since their equivalent width ratios are less sensitive to the abundance of elements. 
For cooler stars however, it is conspicuous the occurrence of neutral helium lines, since strong He\,{\sc ii} 
lines in the optical domain are constrained to stars hotter than B2. As most of the B and Be-star spectra 
in our sample show helium lines in only one ionization state (He\,{\sc i}), we started to study many 
combinations of He\,{\sc i} line ratios and used only those which showed a less dependence on the helium 
abundance. Fixing \logg~ values, we calculated synthetic line profiles for different temperatures from 
15000 to 30000 K and helium number abundances He/H ratios from 0.001 to 0.3. Figures 3 and 4 shows 
several equivalent width ratios of He\,{\sc i} lines against effective temperature and 
He/H abundance ratios obtained for \logg = 3.0 and 4.0. The analyses performed with neutral helium lines 
(He\,{\sc i}$\lambda\lambda$ 4009, 4026, 4121, 4144, 4388, 4438, 4471 and 4922) show that only a few 
combinations of them (He\,{\sc i}4922/He\,{\sc i}4026, He\,{\sc i}4438/He\,{\sc i}4144, 
He\,{\sc i}4438/He\,{\sc i}4026, He\,{\sc i}4438/He\,{\sc i}4009, He\,{\sc i}4144/He\,{\sc i}4121) are 
less sensitive to the helium content (Figures 3 and 4). These less dependent ratios are characterized to be ratios of singlets and
with oscillator strengths ratios near to unity. Triplets are discarded since they depend in a major 
degree on abundance as their statistical weights are higher than for singlet transitions. On the other 
hand transitions with higher oscillator strength values produce profiles with deeper absorptions and more 
dependent on abundance. 
 Once established the final photospheric parameters, we assume that the properties of the stellar core 
are not strongly modified by the rotation, which enables us to determine bolometric luminosities $L$, 
masses $M$ and ages $t$ through interpolation in the evolutionary tracks of non-rotating stars calculated 
by Schaller et al. (1992). In this procedure we used a similar geometrical construction proposed in Myakutin \&
Piskunov (1995) enclosuring the desired solution in a plane $(x,y)$ (where in this case x stands for \teff and 
y for \logg) among two isochrones with ages $\tau_{1}$ and $\tau_{2}$ and two evolutionary tracks with masses 
$m_{1}$ and $m_{2}$ where the four vertices are given by the pairs $(x_{11},y_{11})$, $(x_{12},y_{12})$, 
$(x_{21},y_{21})$ and $(x_{22},y_{22})$. The gravity values in the evolutionary tracks were calculated from 
the relation:

\begin{eqnarray}
\log g = \log {{M}\over{M_{\odot}}} - \log {{L}\over{L_{\odot}}} + 4 \log T_{eff} - 10.6113
\end{eqnarray} 

which arises directly from the Stefan-Boltzmann law. In this system we seek for a solution pair (x,y) contained 
in a straight line that intercepts the tracks $m_{1}$ and $m_{2}$ in the coordinates $(x_{1s},y_{1s})$ and 
$(x_{2s},y_{2s})$. This way we deals with a system of four equations to solve:

\begin{equation}
{{y_{12}-y_{11}}\over{x_{12}-x_{11}}} = {{y_{1s}-y_{11}}\over{x_{1s}-x_{11}}}
\end{equation}

\begin{equation}
{{y_{22}-y_{21}}\over{x_{22}-x_{21}}} = {{y_{2s}-y_{21}}\over{x_{2s}-x_{21}}}
\end{equation}

\begin{equation}
{{y_{1s}-y_{2s}}\over{x_{1s}-x_{2s}}} = {{y_{1s}-y}\over{x_{1s}-x}}
\end{equation}

\begin{equation}
{{(y_{1s}-y_{11})^{2} + (x_{1s}-x_{11})^{2} }\over{(y_{12}-y_{11})^{2} + (x_{12}-x_{11})^{2} }} =  {{(y_{2s}-y_{21})^{2} + (x_{2s}-x_{21})^{2} }\over{(y_{22}-y_{21})^{2} + (x_{22}-x_{21})^{2} }}
\end{equation}

\begin{figure}
\centerline{\epsfxsize= 4.5cm \epsfbox{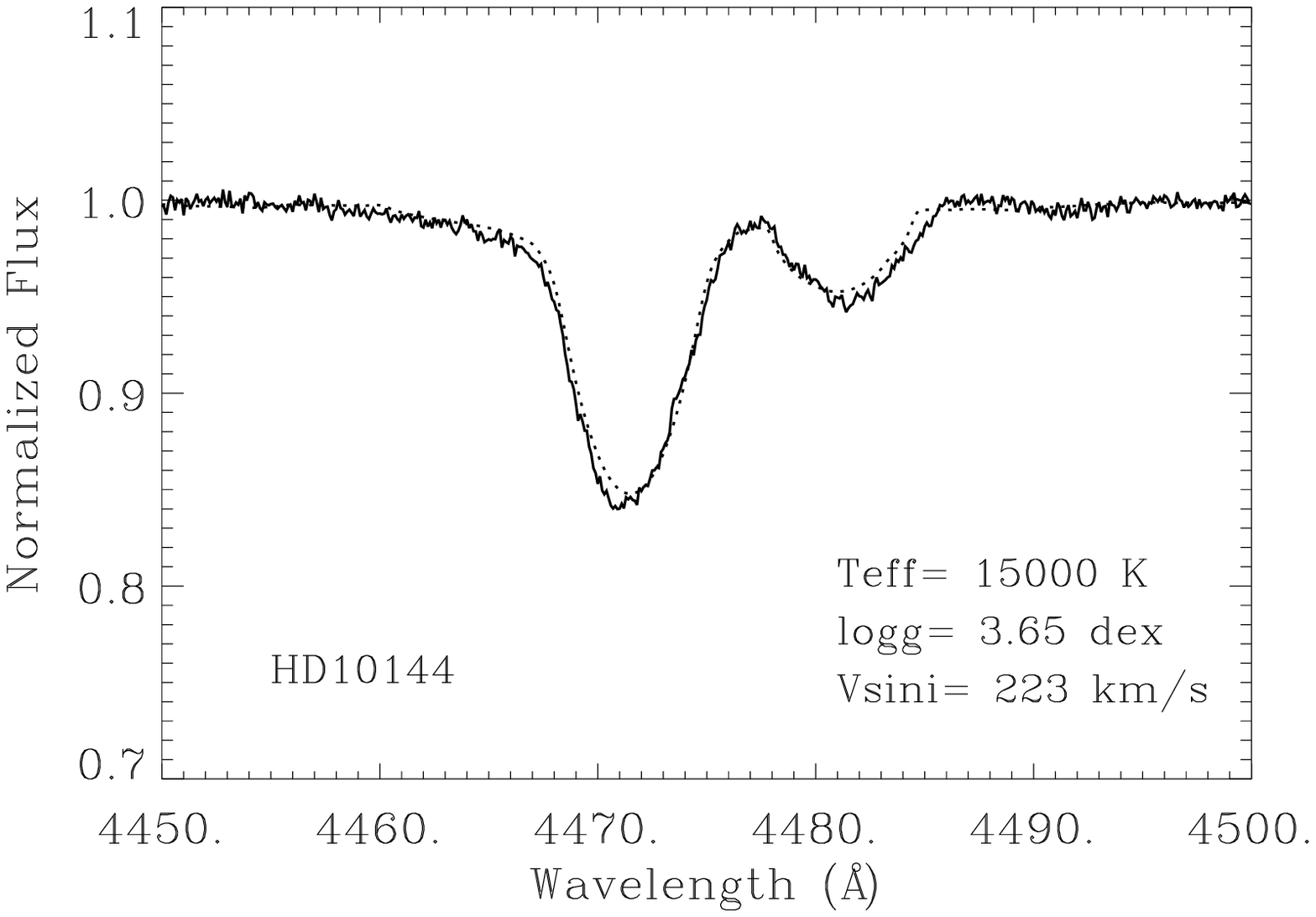} \epsfxsize= 4.5cm \epsfbox{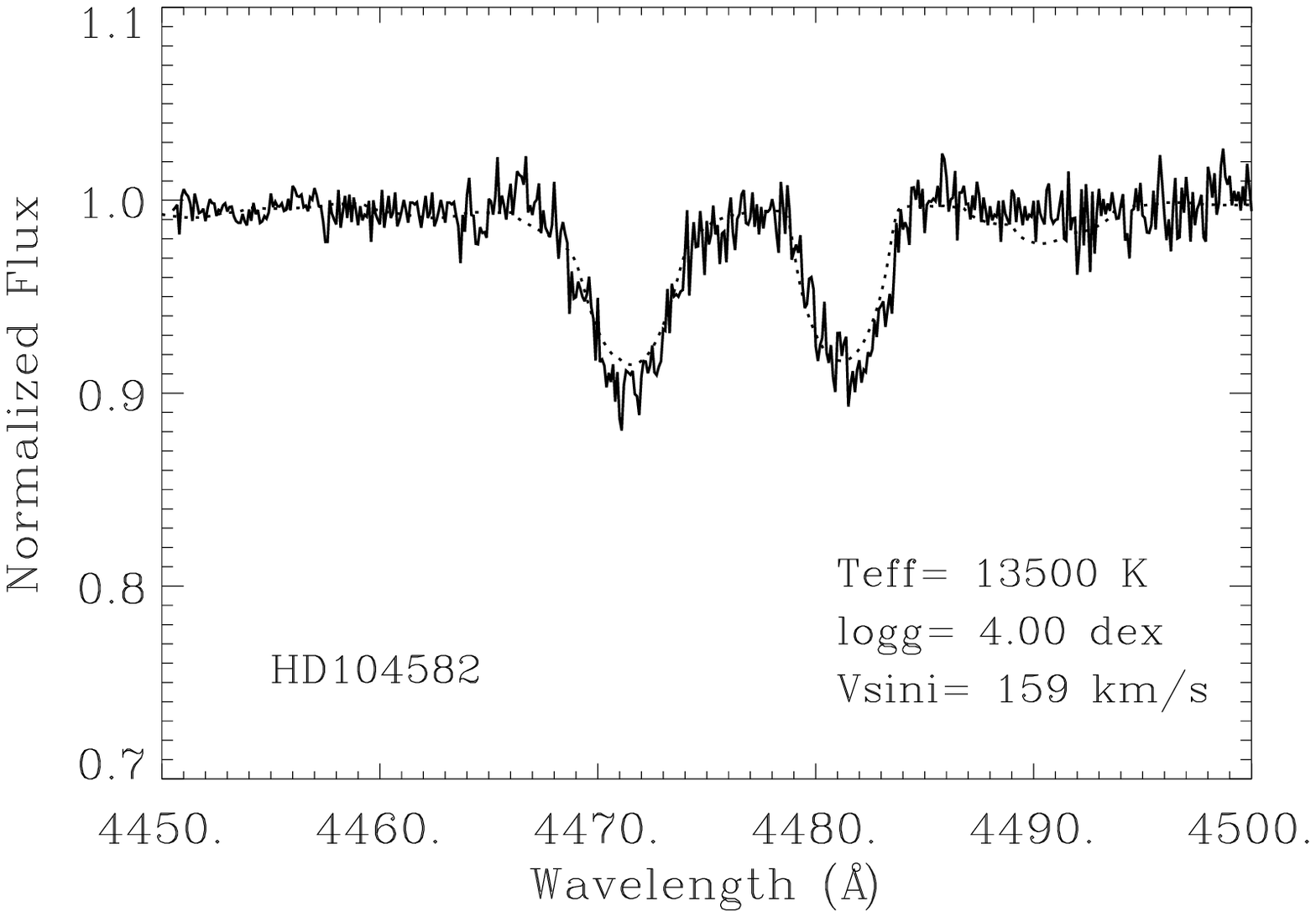}}
\centerline{\epsfxsize= 4.5cm \epsfbox{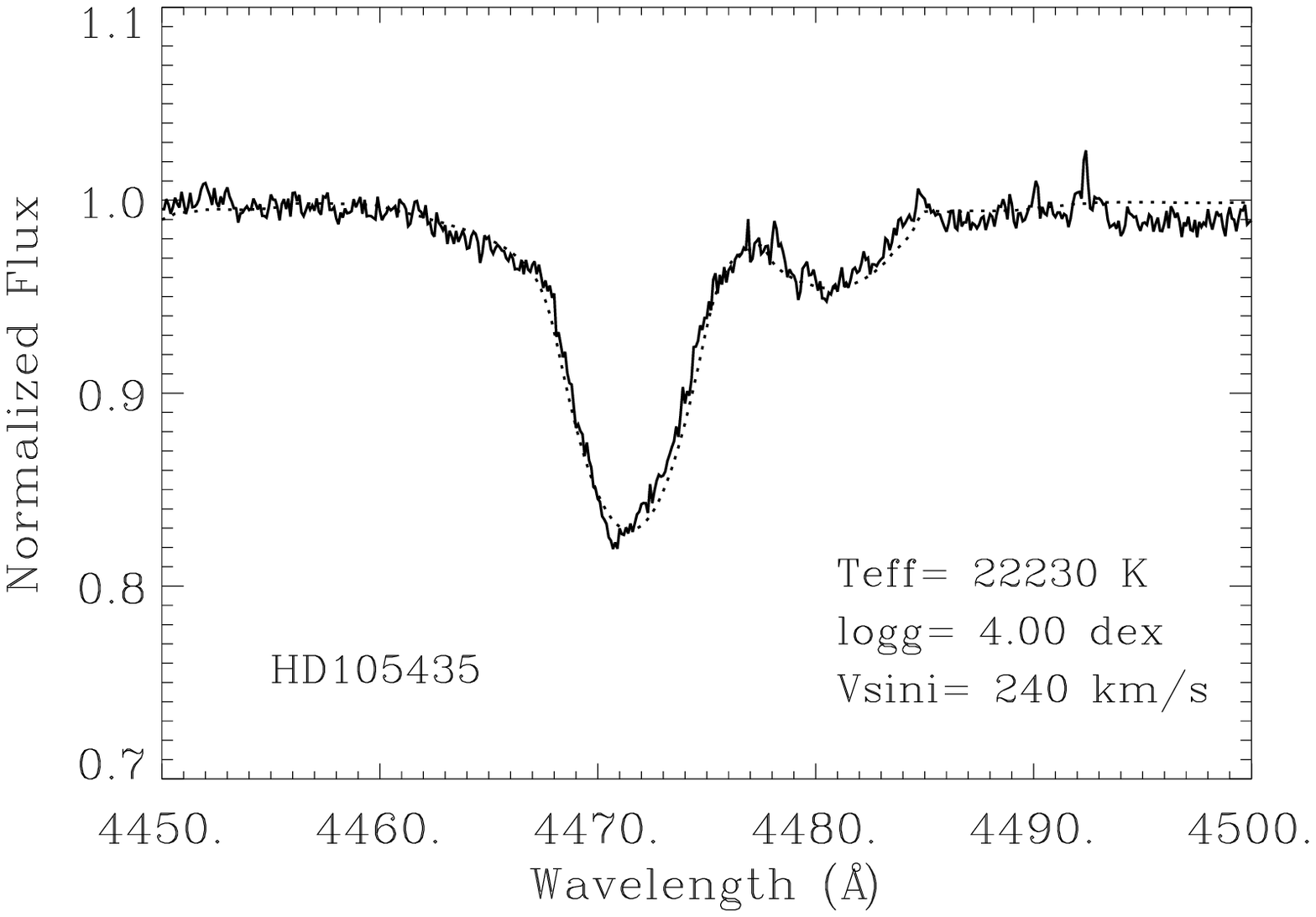} \epsfxsize= 4.5cm \epsfbox{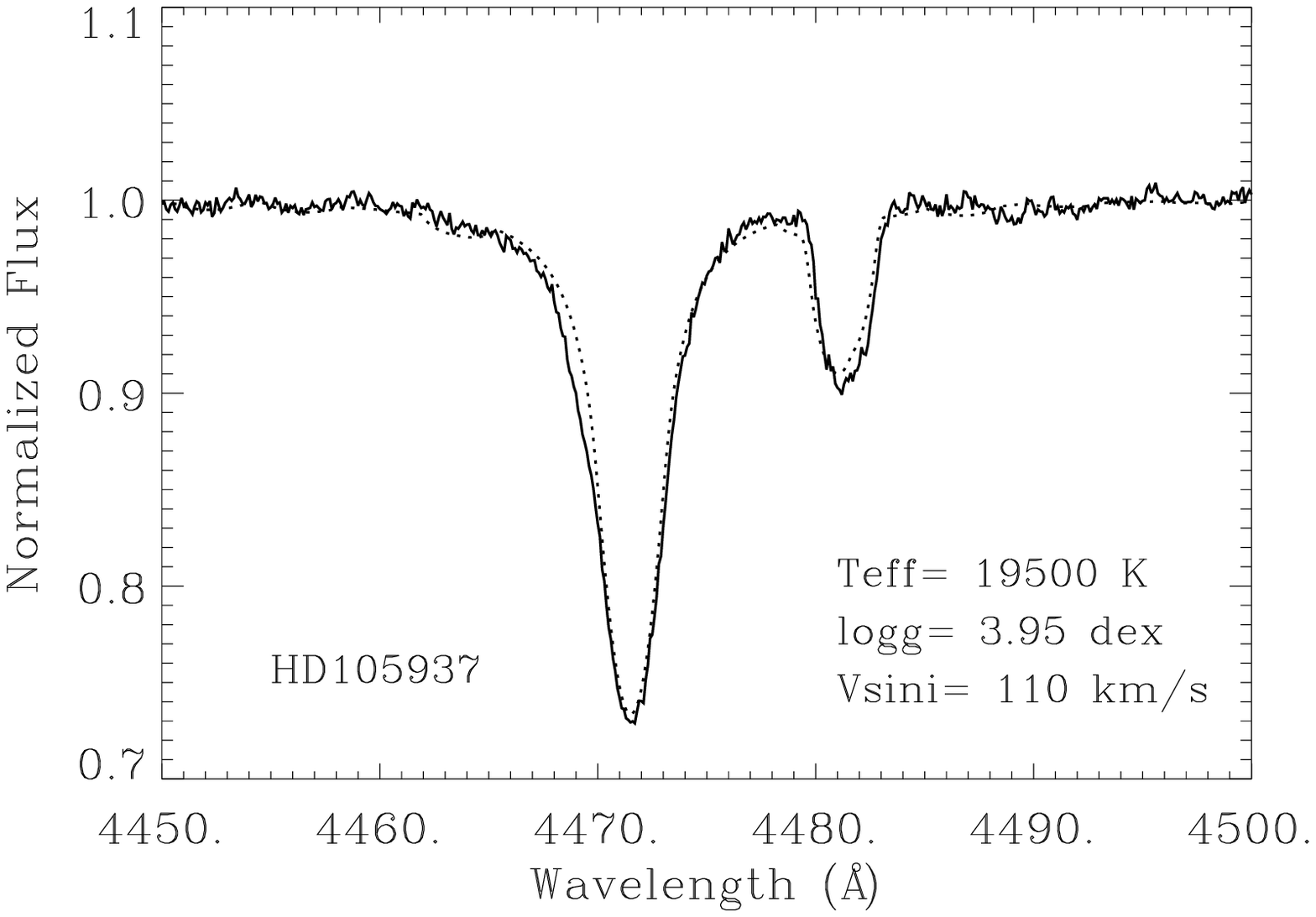}}
\centerline{\epsfxsize= 4.5cm \epsfbox{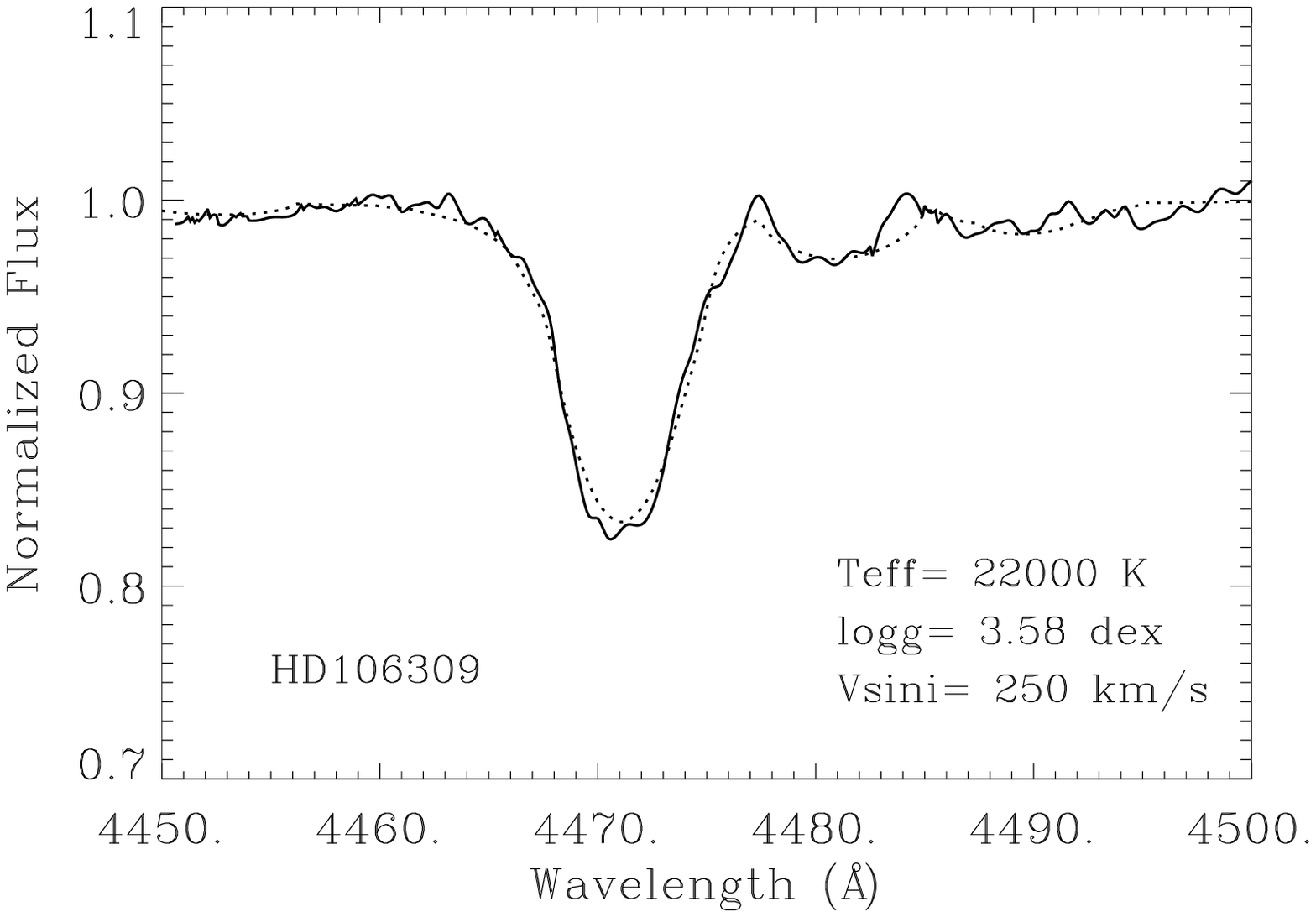} \epsfxsize= 4.5cm \epsfbox{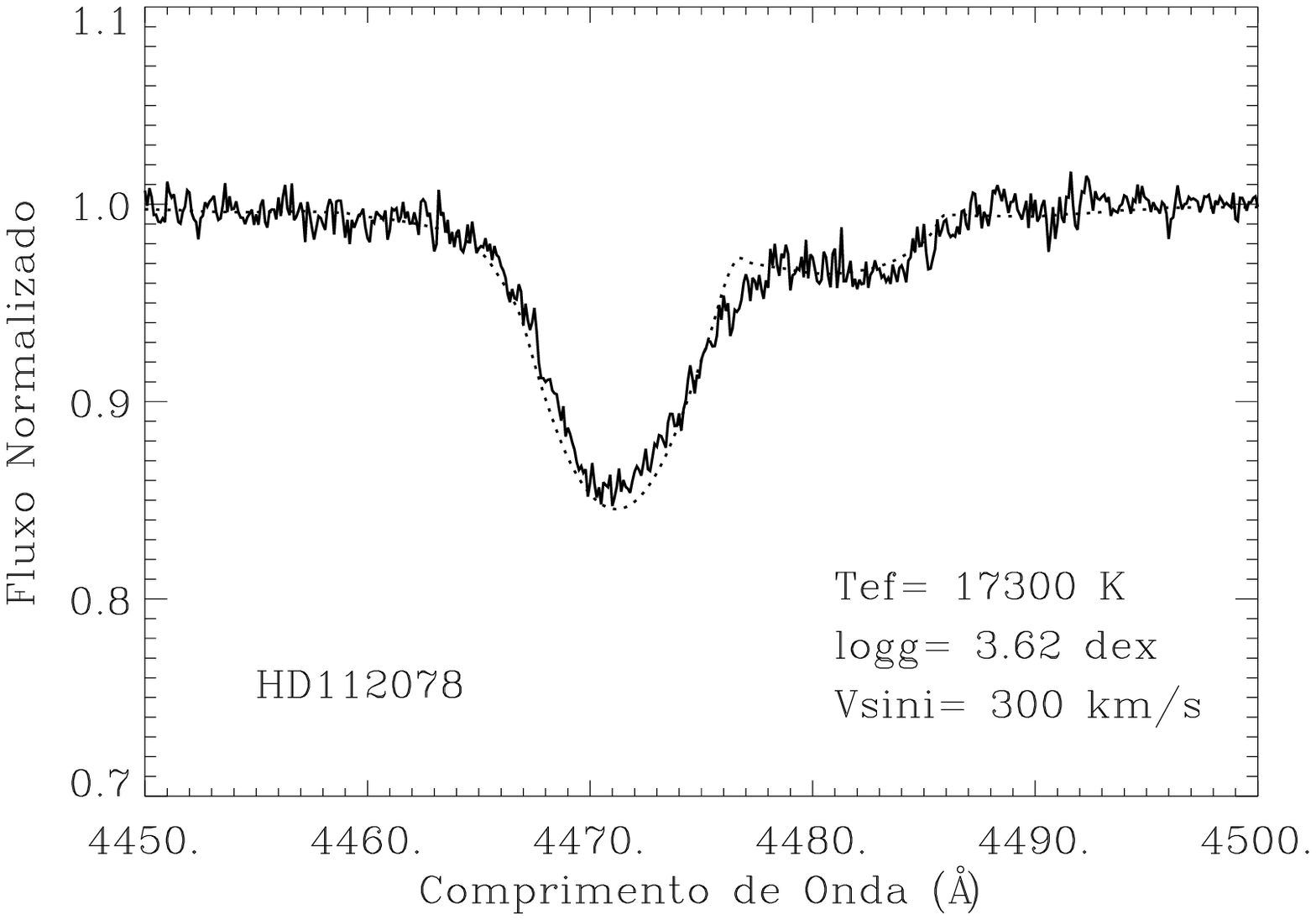}}
\centerline{\epsfxsize= 4.5cm \epsfbox{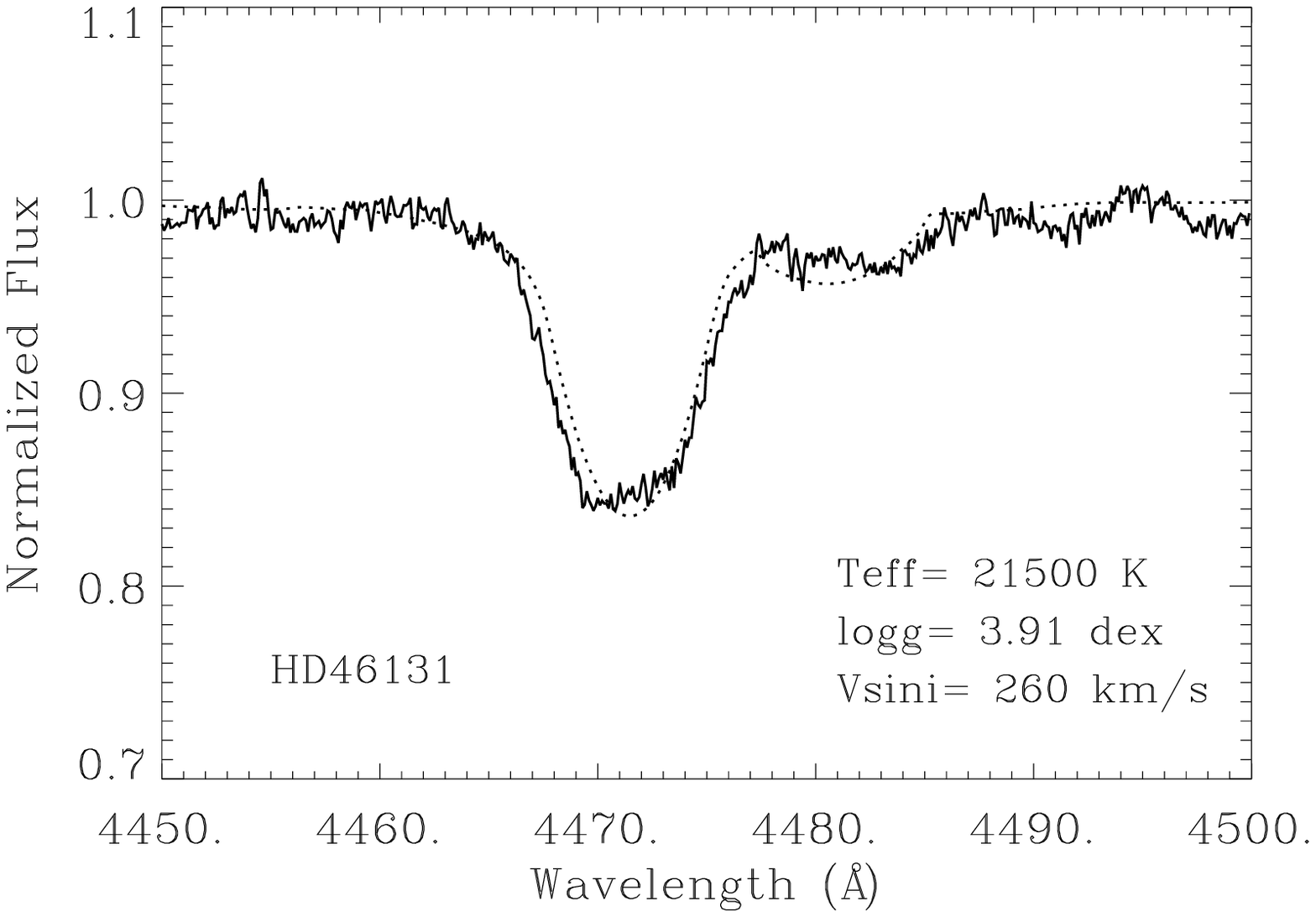} \epsfxsize= 4.5cm \epsfbox{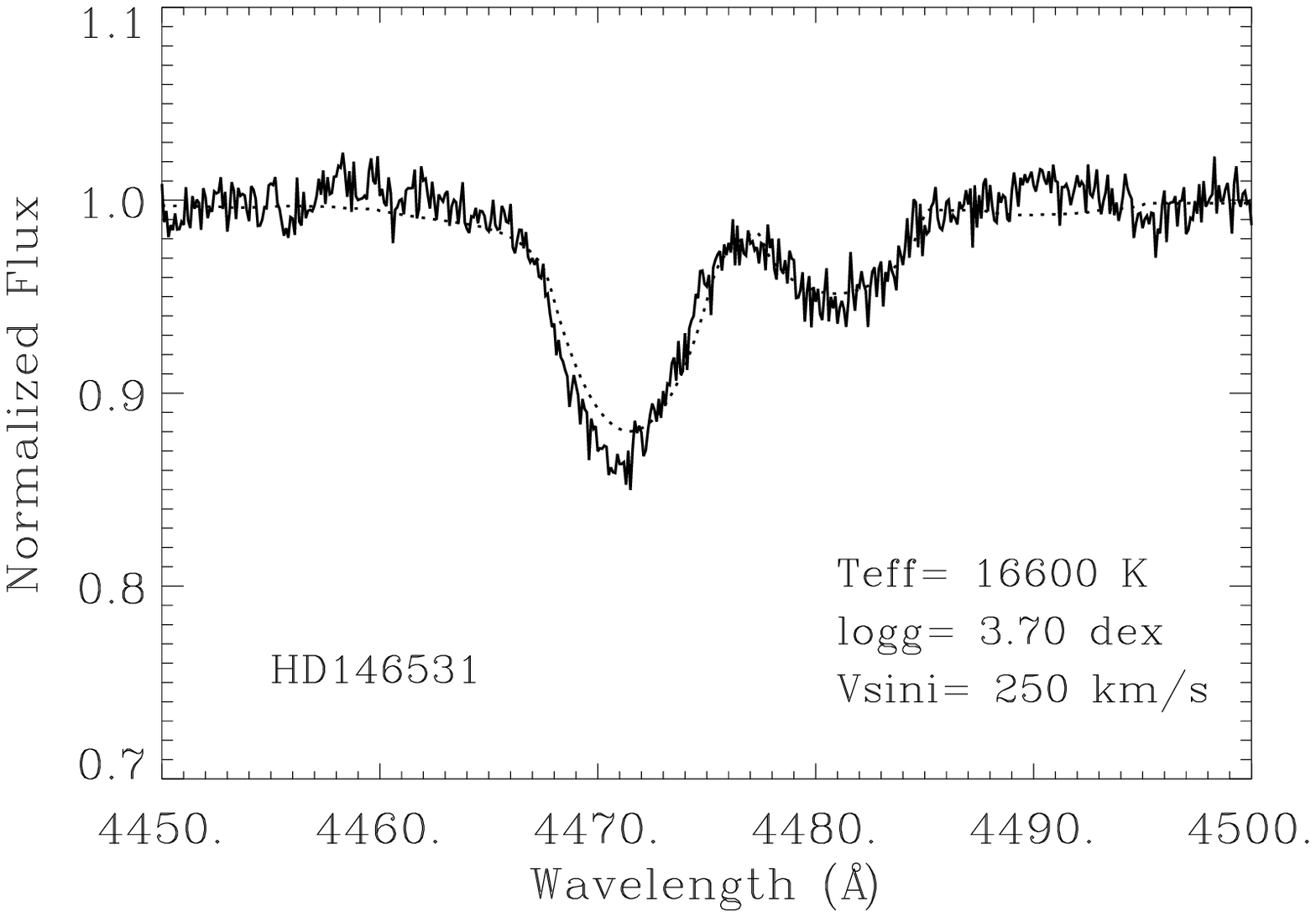}}
\vskip 0.cm
\caption{Sample of Be star spectra fitted with the Nelder \& Mead algorithm (AMOEBA)
in the spectral range of He\,{\sc i}$\lambda$4471\,{\AA} and 
Mg\,{\sc ii}$\lambda$4481\,{\AA} line profiles.}
\label{fig1}
\end{figure}

\begin{figure}
\centerline{\psfig{file=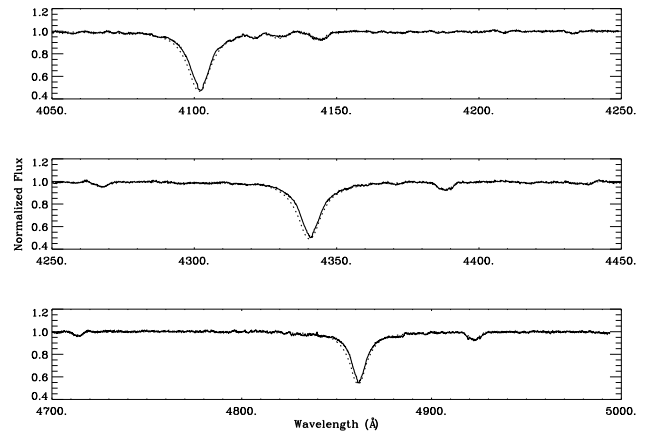,width=8.7truecm,height=6truecm}}
\caption{Comparison of model spectrum (dotted line) with the observed 
photospheric Balmer line profile (full lines) of the Be star Achernar. Upper 
panel: H$\beta$ line; middle panel: H$\gamma$ line; Lower panel: H$\delta$}
\label{fig2}
\end{figure}

\begin{figure}
\centerline{\epsfxsize= 4.5cm \epsfbox{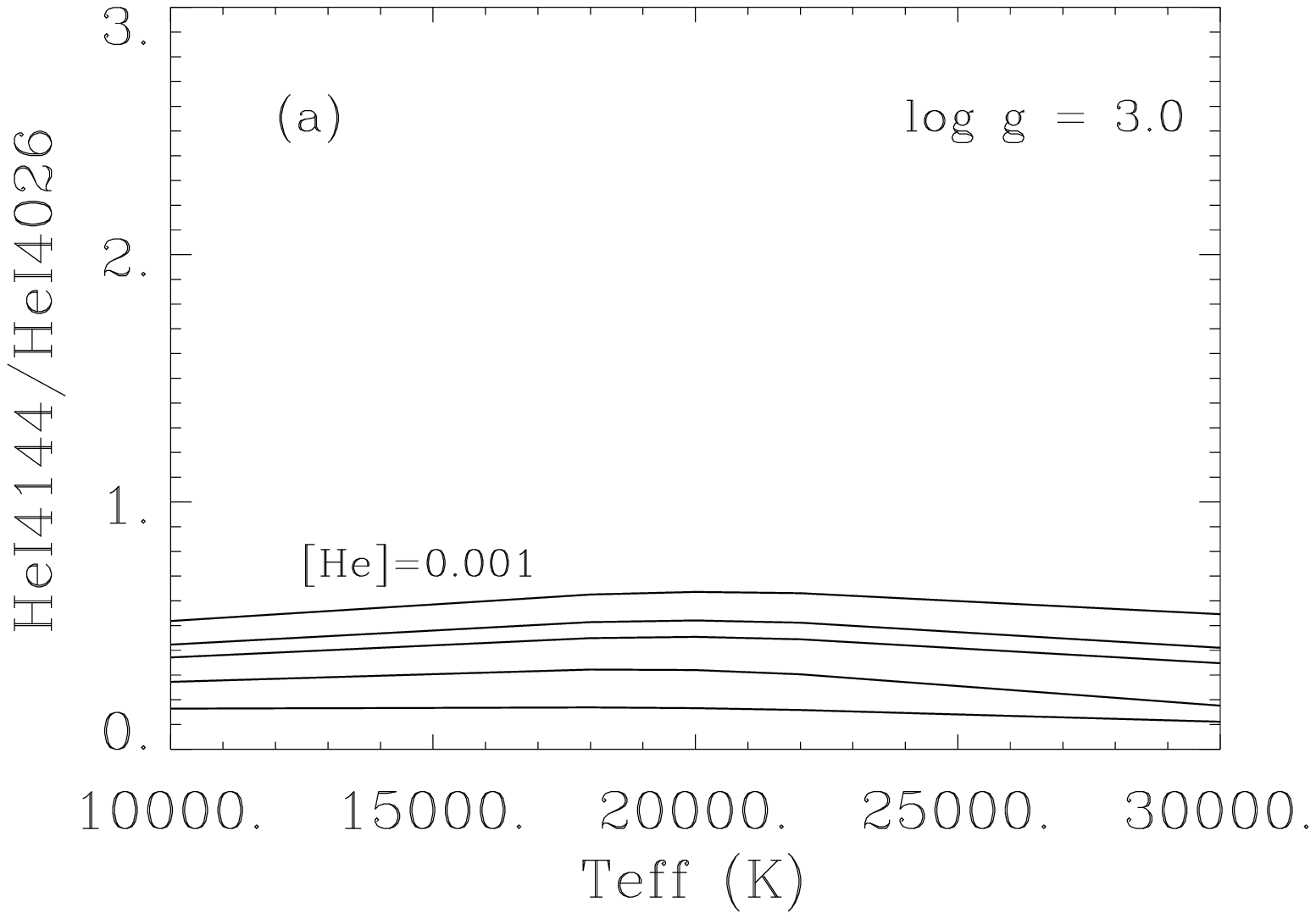} \epsfxsize= 4.5cm \epsfbox{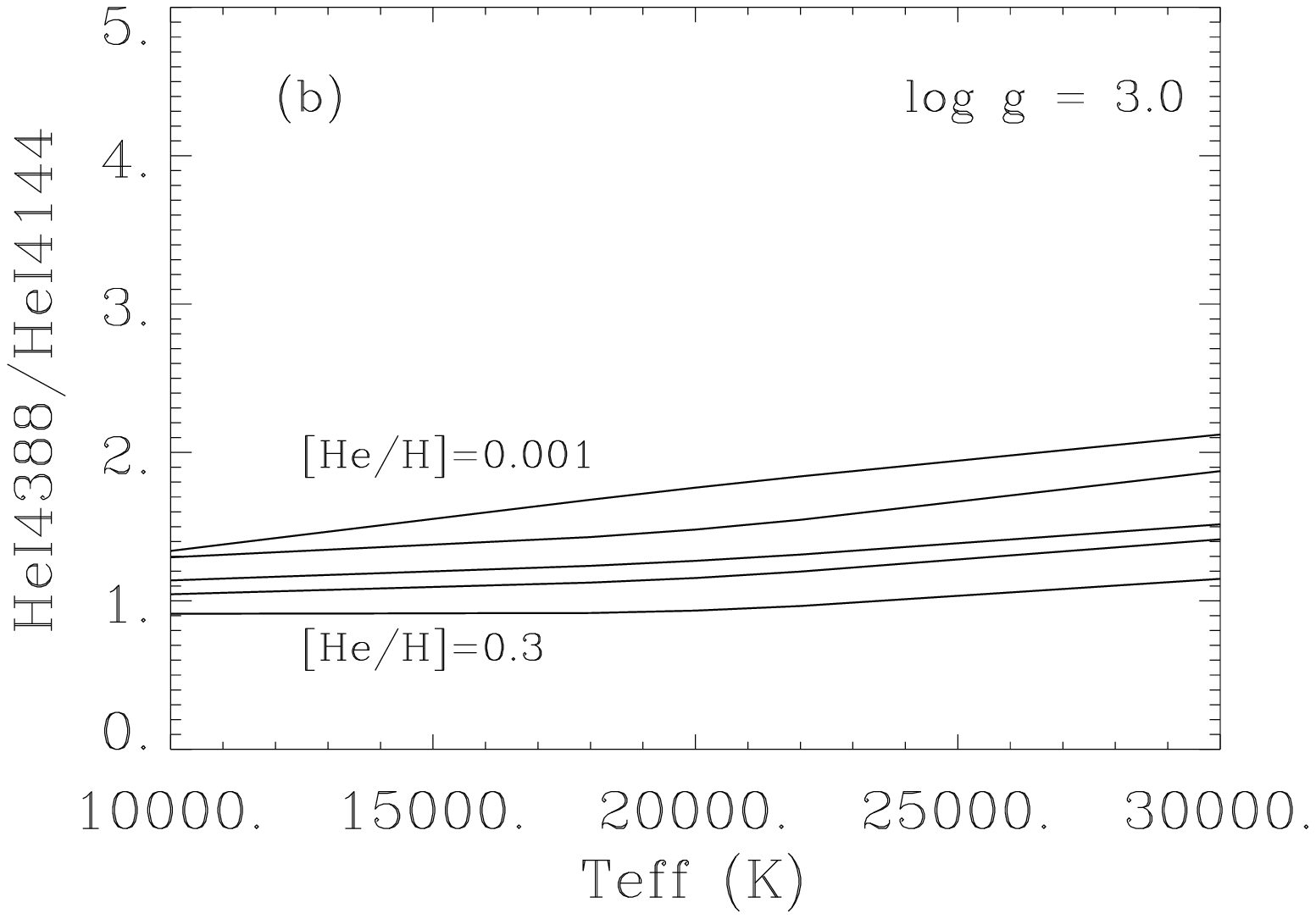}}
\centerline{\epsfxsize= 4.5cm \epsfbox{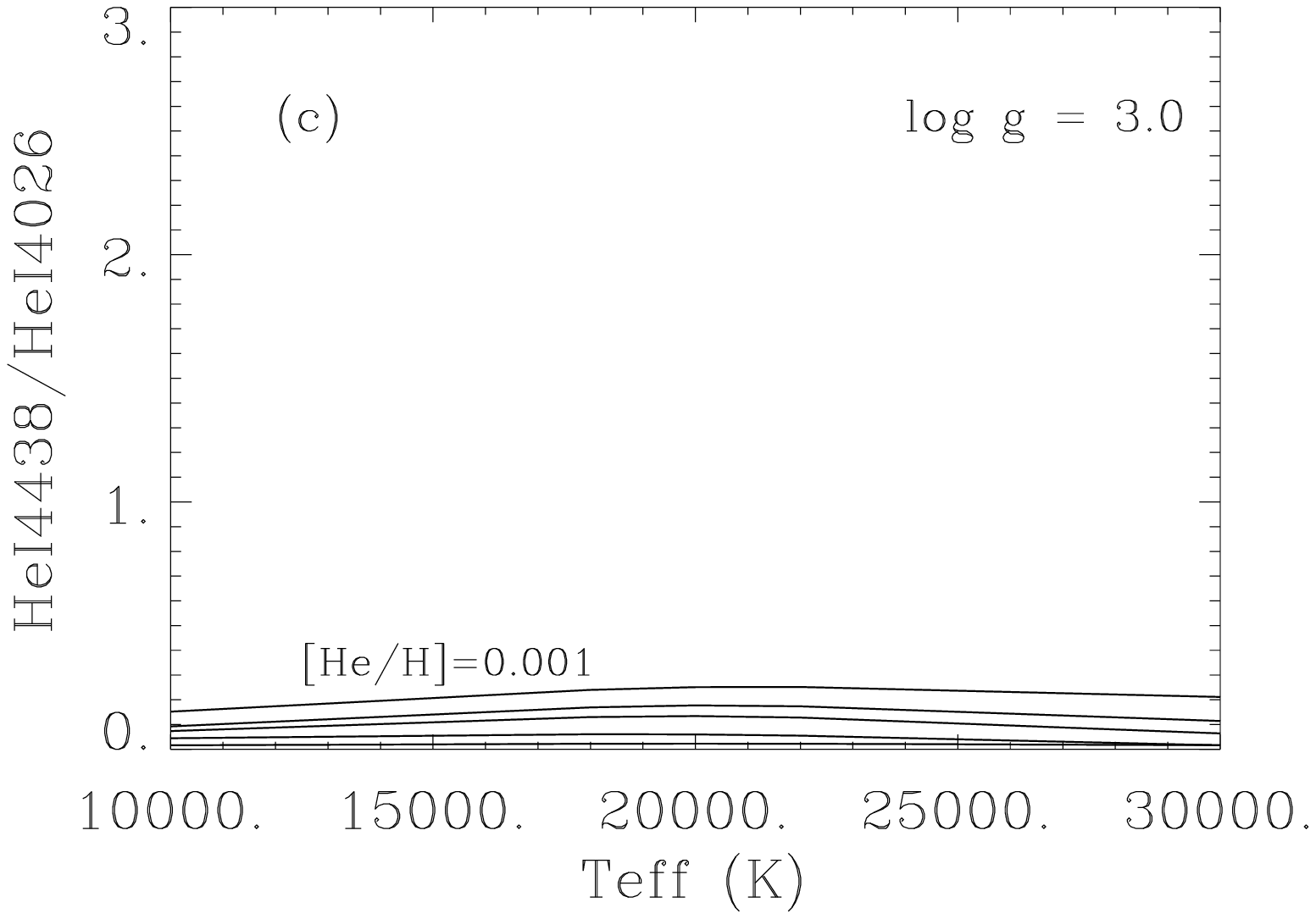} \epsfxsize= 4.5cm \epsfbox{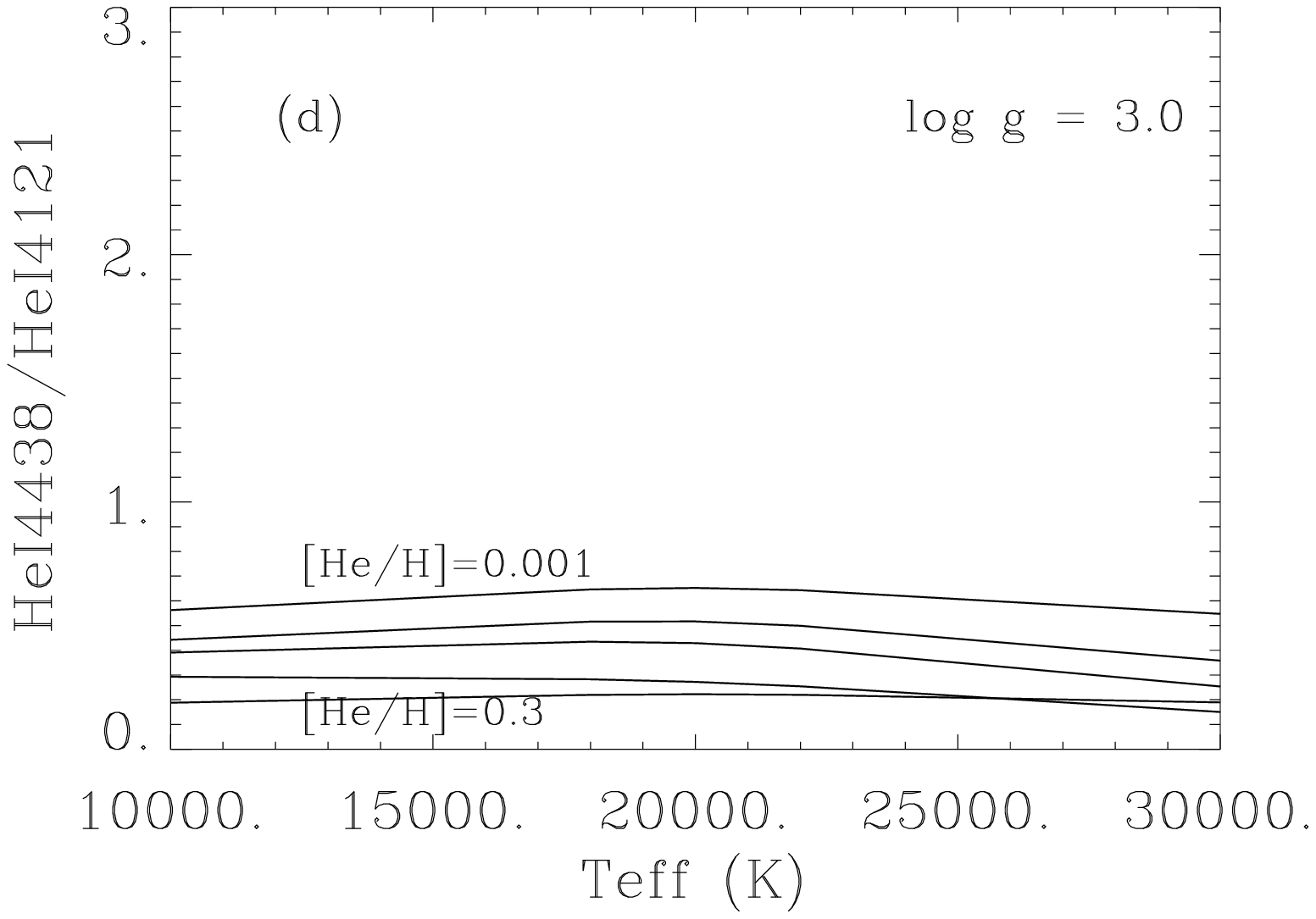}}
\centerline{\epsfxsize= 4.5cm \epsfbox{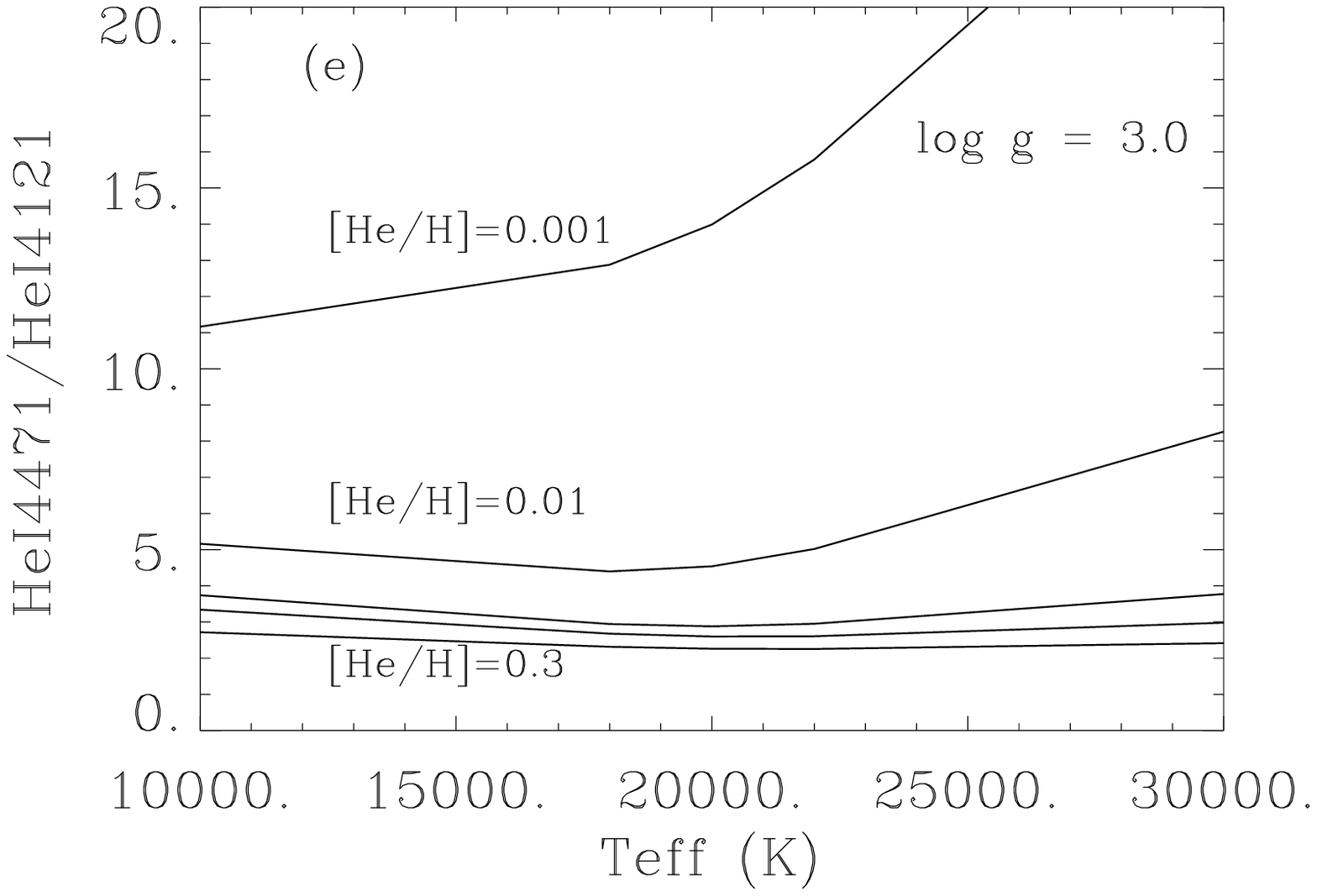} \epsfxsize= 4.5cm \epsfbox{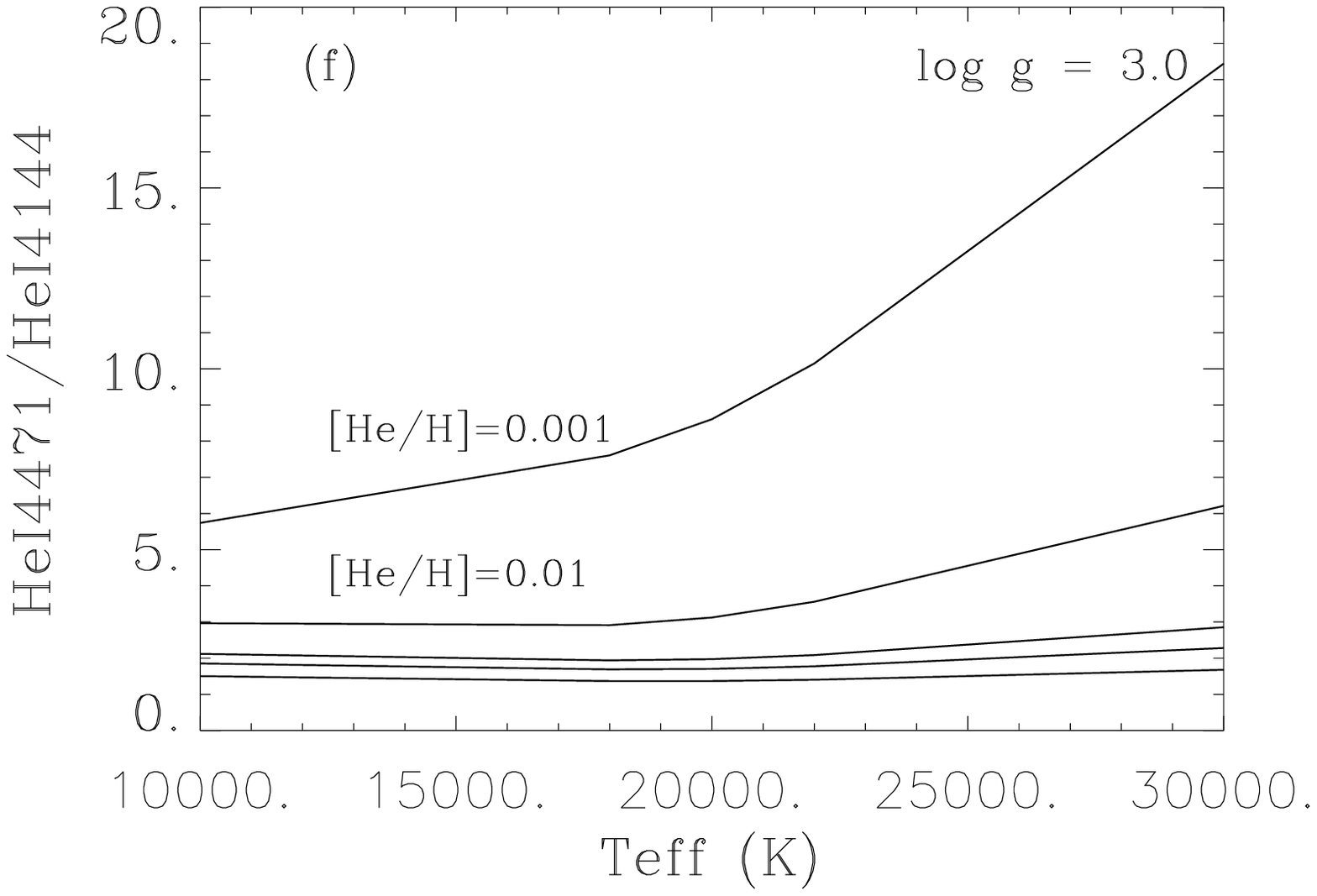}}
\vskip 0.cm
\caption{He\,{\sc i} line equivalent width ratios as function of temperature and abundance, fixing
\logg$=3.0$ dex; (a) equivalent width ratios of He\,{\sc i} 4144/He\,{\sc i} 4026  and (b) He\,{\sc i} 
4388/He\,{\sc i} 4144 do present less dependence with helium content; The same occurs in the case of ratios 
He\,{\sc i} 4438/He\,{\sc i} 4026 (c) and He\,{\sc i} 4438/He\,{\sc i} 4121 (d); In (e) we see the case of 
He\,{\sc i}4471/He\,{\sc i} 4121 and (f) He\,{\sc i} 4471/He\,{\sc i} 4144, where the abundance dependence is higher.}
\label{fig3}
\end{figure}

\begin{figure}
\centerline{\epsfxsize= 4.5cm \epsfbox{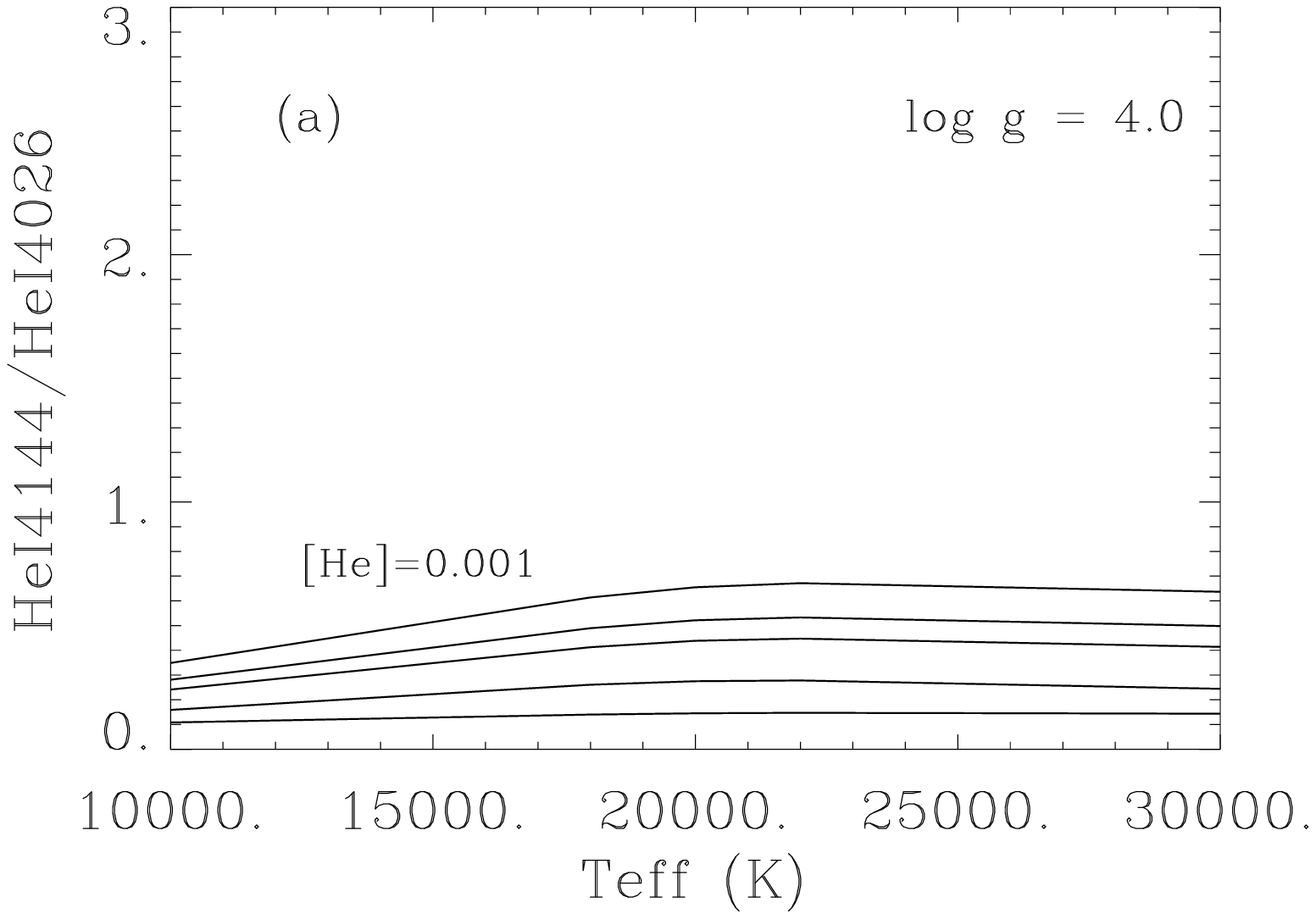} \epsfxsize= 4.5cm \epsfbox{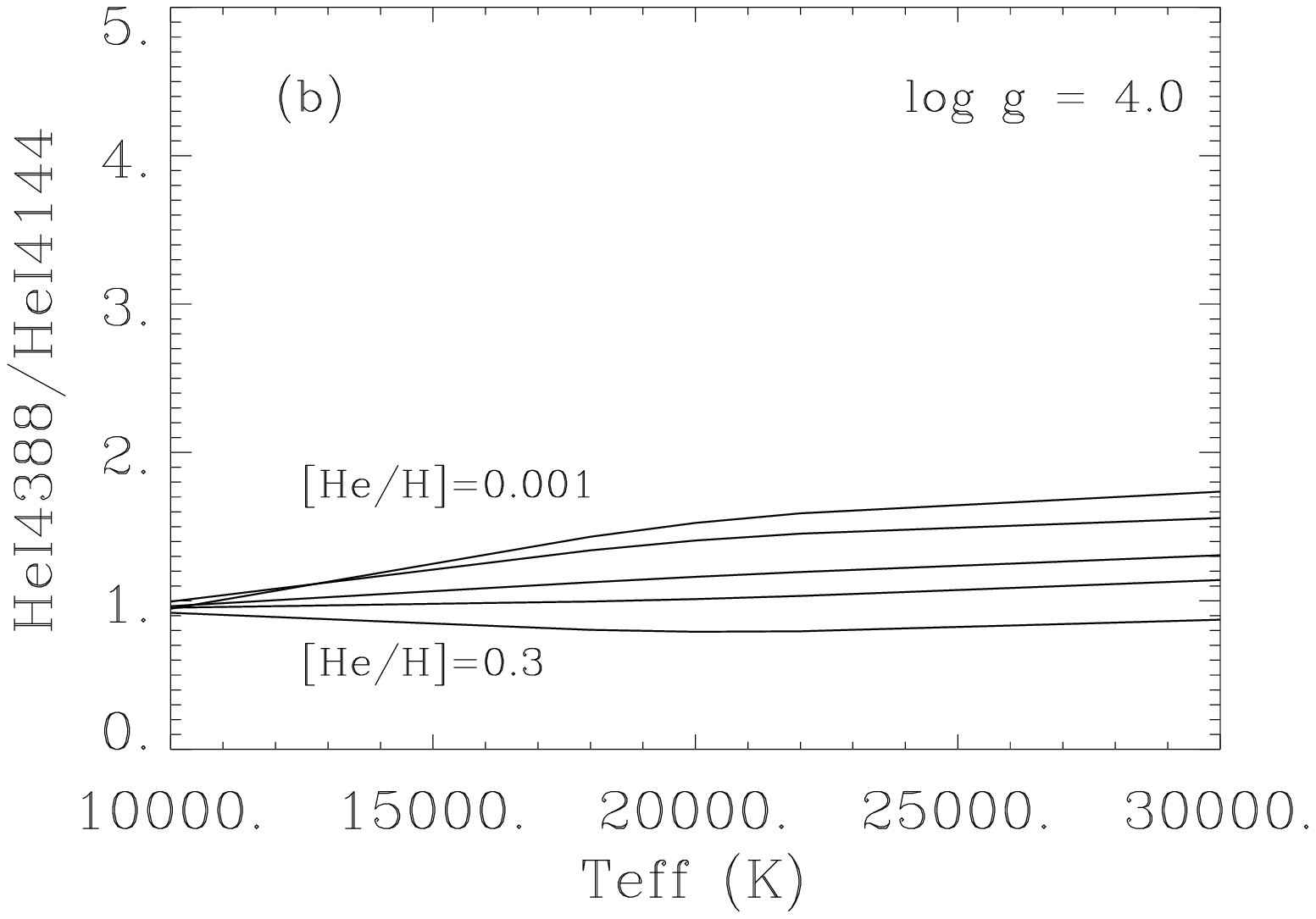}}
\centerline{\epsfxsize= 4.5cm \epsfbox{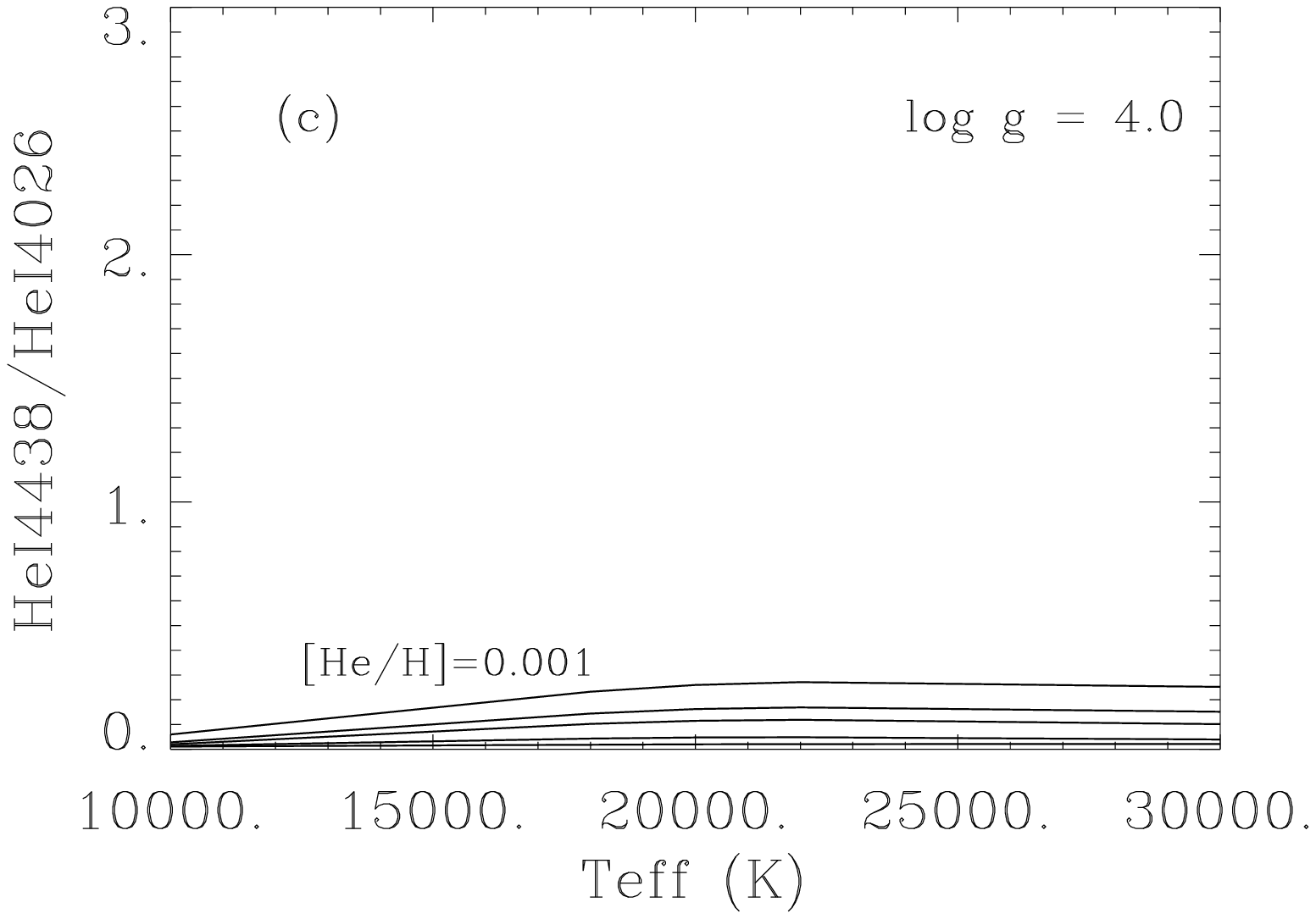} \epsfxsize= 4.5cm \epsfbox{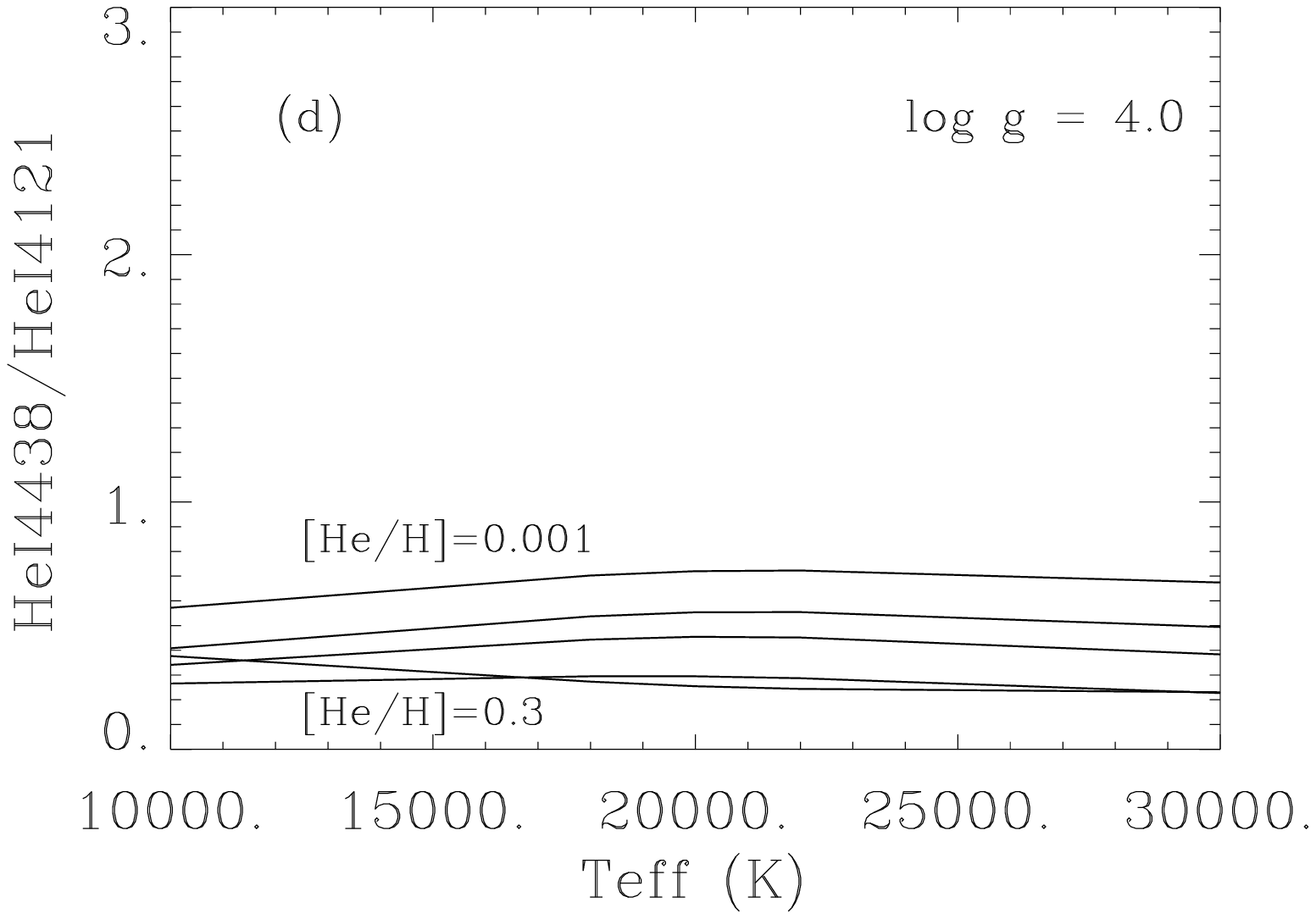}}
\centerline{\epsfxsize= 4.5cm \epsfbox{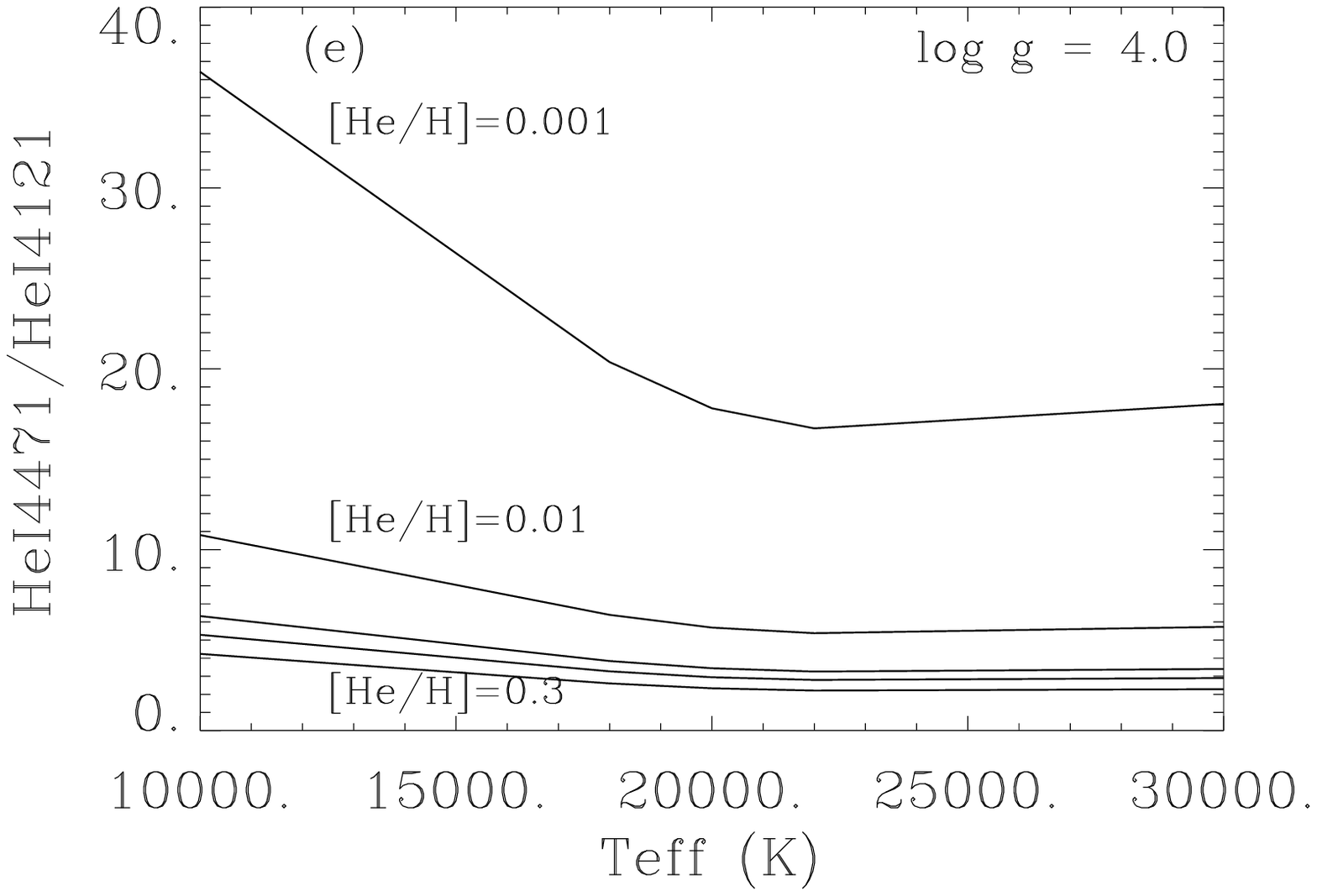} \epsfxsize= 4.5cm \epsfbox{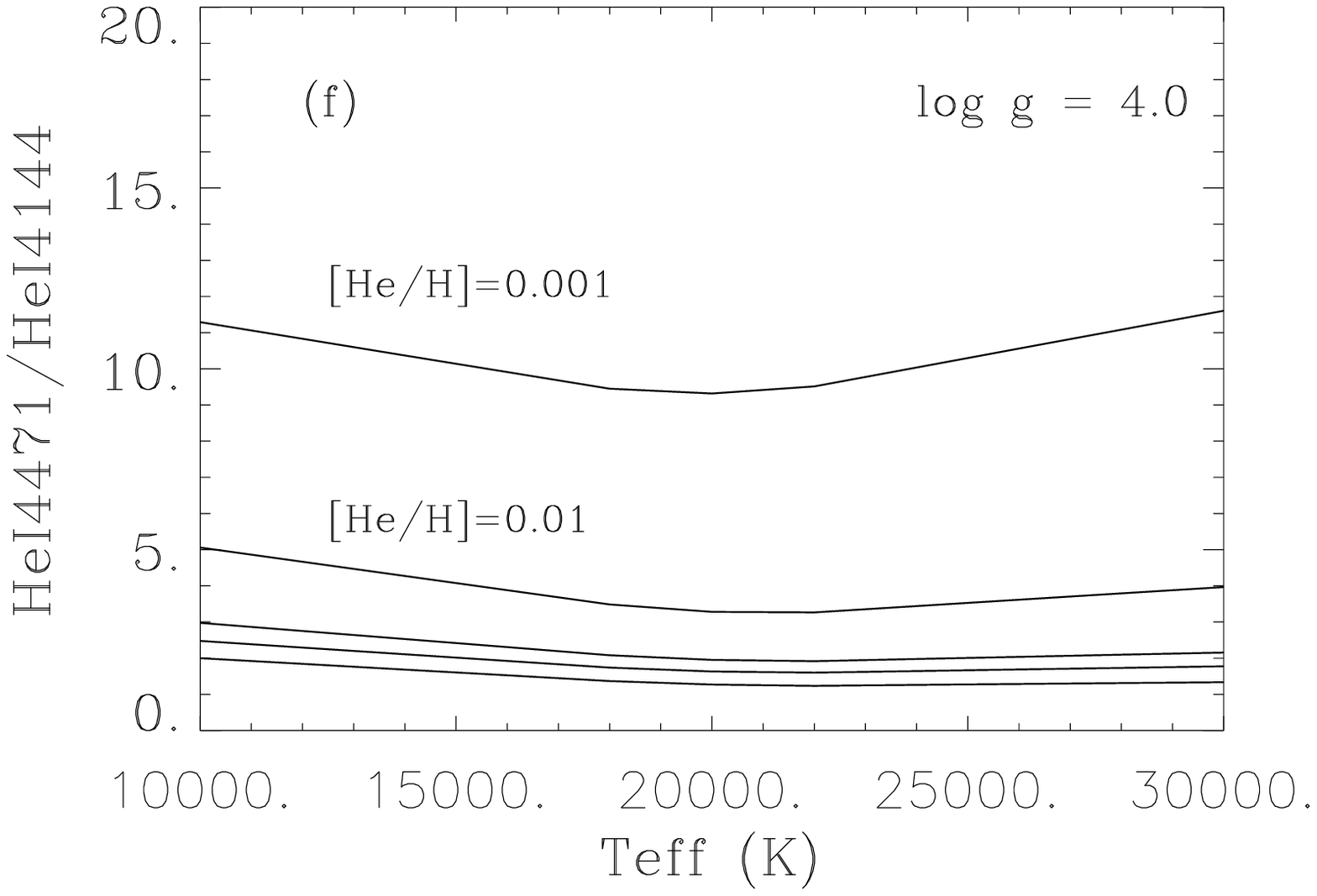}}
\vskip 0.cm
\caption{Same as figure 3, but for \logg$=4.0$ dex.}
\label{fig4}
\end{figure}

 The fundamental parameters determined in this work are presented in Table 2. 
Columns 1, 2, 3 and 4 stands for the object's name (HD code), projected rotational 
velocity (\vsini in $\rm km.s^{-1}$), effective temperature (\teff in K) and
superficial gravity (\logg in dex) respectively. By interpolating 
in the Schaller's et al. (1992) theoretical evolutionary tracks (taking $Z=0.02$) 
using the effective temperatures and gravities determined from the analysis of 
line spectra we obtain estimates of their logarithmic ages, bolometric luminosities 
$\log L/L_{\odot}$ and masses $M/M_{\odot}$, represented in columns 5, 6 and 7 
of Table 2 respectively. The spectral types and luminosity classes presented 
in column 8 are taken from the SIMBAD database at the Centre de Donn\'ees de 
Strasbourg (CDS). In column 9 it is provided the spectral classification based
on the photospheric parameters evaluated in this work. In columns 10 and 11
we furnish estimates of deviations in temperature and gravity due to the
effects of rapid rotation. These estimates were inferred, only in the case 
of Be stars, from the direct fitting of NLTE profiles affected by gravity 
darkening to the observed spectra. Column 12 shows the fractional age 
$\tau/\tau_{\rm M.S.}$ only for Be stars. The positions of Be stars 
in the HR diagram are shown in Figure 5, whereas the evolutionary 
stages, corresponding to the entries in column 12, as a function of stellar 
masses are given in Figure 6, together with histograms of Be frequency counts
for each specific mass range.

\begin{figure}
\centerline{\psfig{file=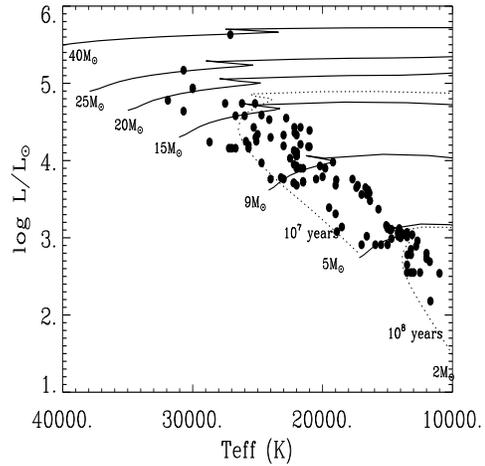,width=7truecm,height=7truecm}}
\caption{Location in the HR diagram ($\log L/L_{\odot}$,\teff) of all
observed Be stars neglecting the effects of high rotation. The evolutionary tracks 
(full lines) and isochrones (dashed lines) are from Schaller et al. (1992)} 
\label{fig5}
\end{figure}

\begin{figure}
\centerline{\epsfxsize= 5.cm \epsfbox{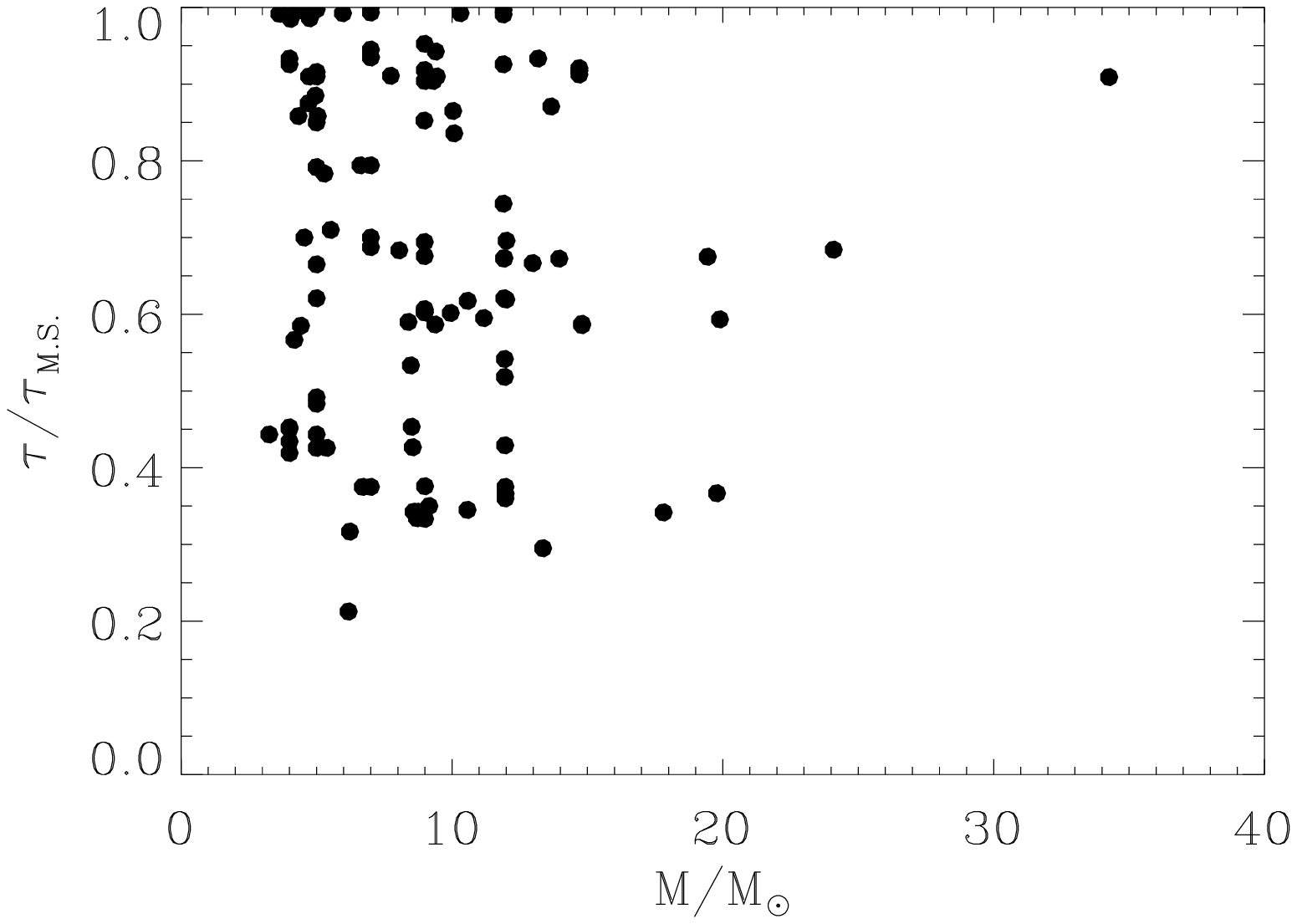} \epsfxsize= 5.cm \epsfbox{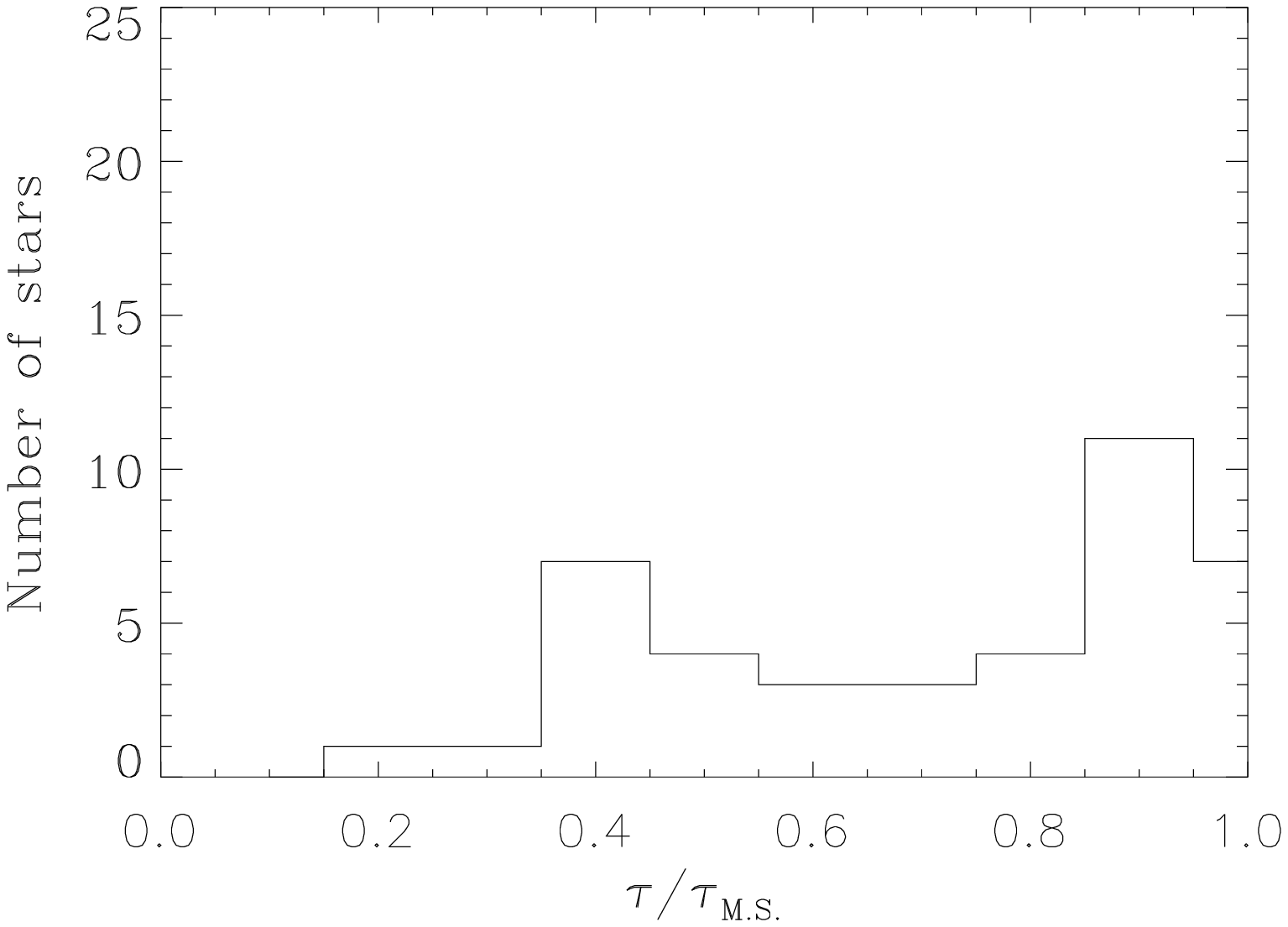}}
\centerline{\epsfxsize= 5.cm \epsfbox{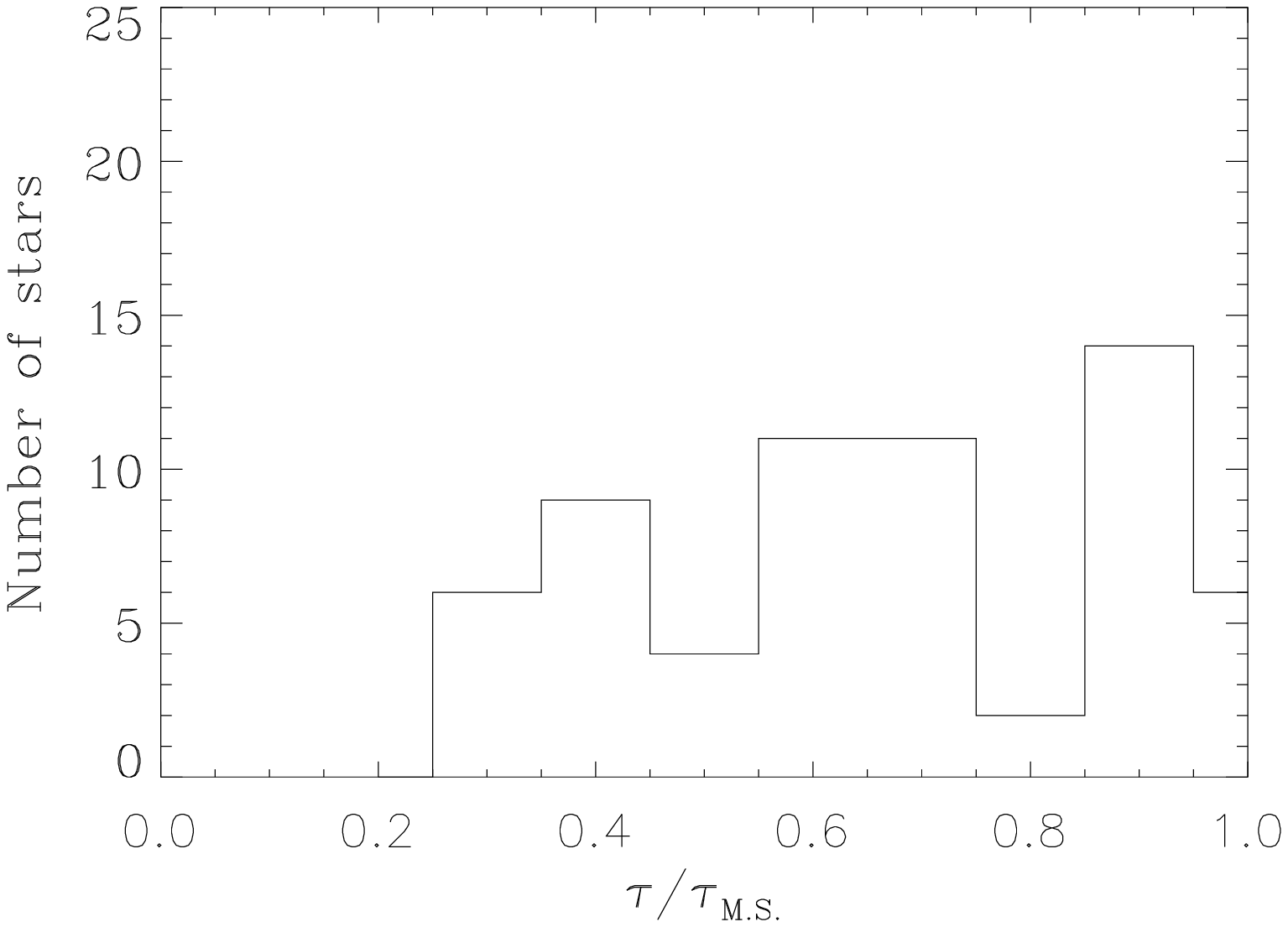} \epsfxsize= 5.cm \epsfbox{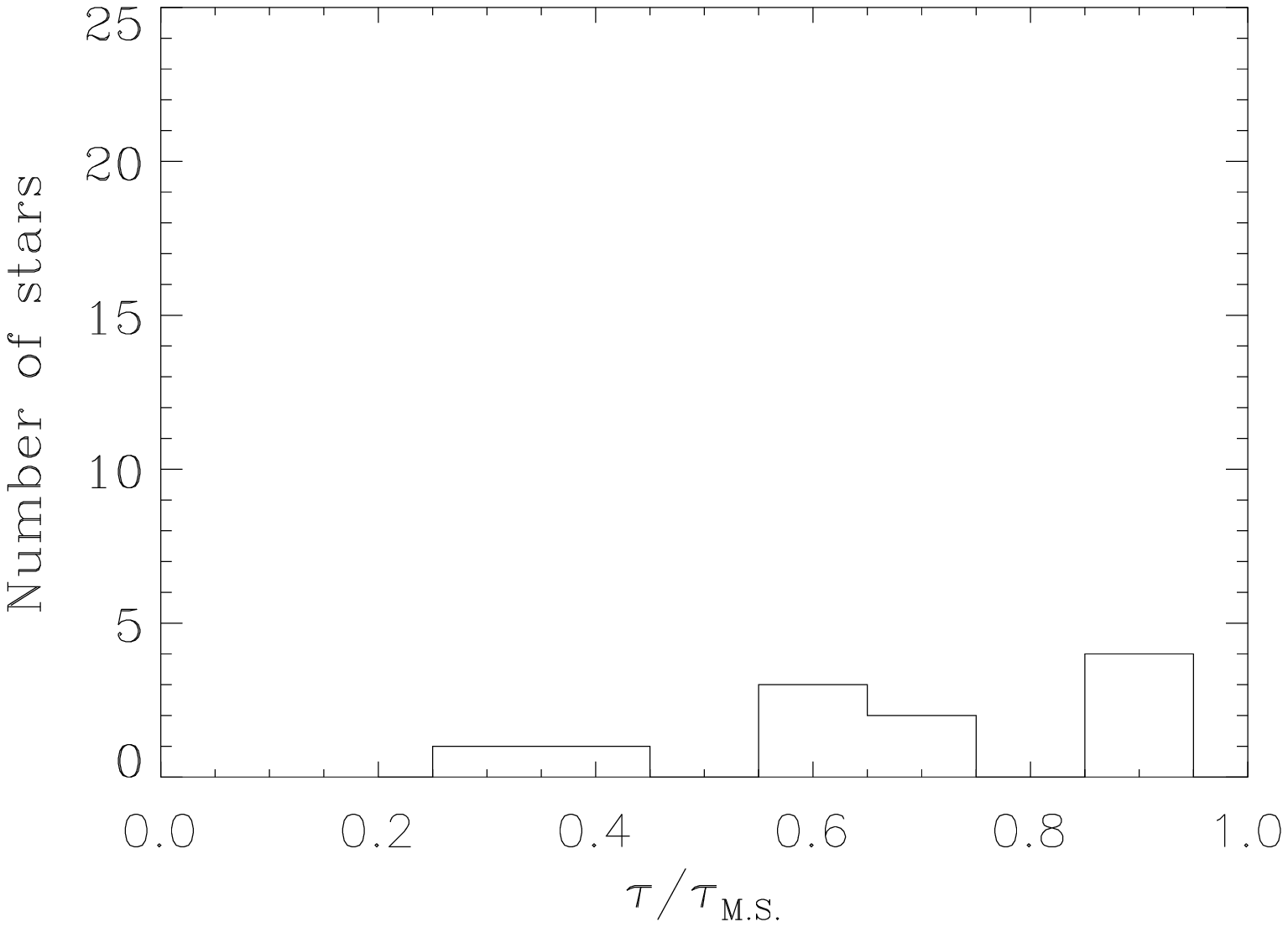}}
\vskip 0.cm
\caption{(up left) $\tau/\tau_{SP}$ diagram as a function of mass for 114 field Be stars. (up right) Frequency 
counts of Be stars of low mass ($< 7 M_{\odot}$) as a function of  evolutionary stage in the mean sequence. 
(down left) Frequency counts of Be stars with intermediary mass (between 7 and $14 M_{\odot}$) as a function of 
evolutionary stage. (down right) The same, but for higher masses (above $14 M_{\odot}$).}
\label{fig6}
\end{figure}

\subsection{Discussion}

\noindent In this work we presented our first results of southern field B and Be star's photospheric parameters. The determination of effective temperatures (\teff), superficial 
gravities (\logg) and projected rotation velocities (\vsini) was achieved in three steps. 
In the first one we derived an initial value of \vsini~ from the average of values 
estimated by Fourier Transforms (FT) of all neutral helium line profiles in the 
wavelength domain from 4000 to 5000 \AA~. In the second step we derived \teff~ and 
\logg~ from the ionization equilibria of He, O, N, Si and Balmer equivalent width fits, 
which delimitate various minima in the Kiel diagram. In case of stars with strong 
emissions we adopted the background photospheric fitting procedure described in Chauville 
et al. (2001). We then adopted photometric indices in order to choose a well suitable 
region to apply spectral synthesis. In the last step we used a more accurate fitting 
procedure on conspicuous H and He lines with synthetic NLTE spectra (neglecting gravity 
darkening) to the observed data using the downhill simplex algorithm (AMOEBA) and limited 
the search to the best regions mentioned above. 
As our stellar sample are in average constrained to the solar vicinity, we can in principle consider all of them as possessing the same metallicity values (i.e. solar ones). 
Inspecting figure 6, in the $\tau/\tau_{\rm M.S.}$ x $\rm M/M_{\odot}$ diagram, we see 
clearly a tendency of occurrence of Be stars in the second half of the main sequence 
phase. Dividing this stellar sample in three subgroups, namely those with low mass 
($M < 7 M_{\odot}$), intermediate ($\rm 7 M_{\odot} < M < 14 M_{\odot}$), and higher 
masses ($M > 14 M_{\odot}$), we get the histograms shown in the same figure. According
to that we can have a ``fast visual insight" on the dependence of the occurrence of
the Be phenomenon with the stellar mass. For low massive stars, the Be phenomenon 
starts to occur even in earlier stages in the main sequence, but preferably in the
second half of the main sequence, as they evolve more slowly and have time enough 
for that. As the stellar mass increases, it turns out to be more difficult to find 
Be stars in earlier stages, as they walk faster to the TAMS and this way it is more 
difficult to precise when the star started to display the phenomenon. 

As a general result, without taking into account for rotational effects on the stellar
spectra, B stars seem to display the Be phenomenon in the second half of the main sequence  evolutionary phase. This result implies the existence of some evolutionary change 
that enables the appearance of the Be phenomenon during the MS lifetime and agree with 
those reported by many authors (McSwain \& Gies 2005; Fabregat \& Torrej\'on 2000; 
Meynet \& Maeder 2002; Keller et al. 2001, Lyubimkov 1998). On the other hand, Zorec, 
Fr\'emat \& Cidale (2005) analysed a sample of 97 Be stars using models of rotating stars
and showed that, when rotational effects take place, the Be phenomenon occurs 
elsewere in the main sequence.

In order to constrain statistically the role of mass in the occurrence of the Be 
phenomenon, as well as to confirm the effects of gravity darkening in its appearance
among B stars, it is important to analyse in the future a larger stellar sample.

\section*{Acknowledgments}
RSL and NVL warmly thanks and are in debt to Dr. J. Zorec for his advices as well as to Drs. Y. Fr\'emat, J. Chauville, D. Ballereau,
A.M. Hubert and M. Floquet for fruitful discussions. The authors are also grateful to an anonymous referee for valuable comments
and suggestions which improved this work. This research was supported by FAPESP (Funda\c{c}\~ao de Amparo \`a Pesquisa do Estado de S\~ao Paulo) 
through grants no. 04/14256-8, 00/10029-6 and 1998/10138-8.

\eject

\addtocounter{table}{+0}%
\begin{table*}
\begin{centering}
\rotcaption{Photospheric parameters of target stars.}
\begin{sideways}
\tiny{
\begin{tabular}{l|ccccccllccc}
\hline\hline
Object  &  \vsini (\kps)   &  \teff (K)  &  \logg (dex)  & \lage (yr)  &  \logl &  \mmsun &  Sp.T.(Simbad) &  Sp.T.(this work) & $\Delta$\teff (K) & $\Delta$\logg (dex) & $\tau/\tau_{\rm M.S.}$ \\
\hline\hline
HD	10144	&	$	223	\pm	15	$	&	$	15000	\pm	600	$	&	$	3.60	\pm	0.10	$	&	$	7.99	\pm	0.08	$	&	$	3.14	\pm	0.10	$	&	$	4.9	\pm	0.4	$	&	B3Vpe	&	B6Vpe	&	900	&	0.2	&	0.91	\\
HD	14850	&	$	150	\pm	20	$	&	$	13500	\pm	550	$	&	$	3.60	\pm	0.10	$	&	$	8.05	\pm	0.06	$	&	$	3.01	\pm	0.10	$	&	$	4.75	\pm	0.10	$	&	B8Ve	&	B8Ve	&	350	&	0.1	&	0.99	\\
HD	15371	&	$	10	\pm	8	$	&	$	14700	\pm	400	$	&	$	3.50	\pm	0.10	$	&	$	7.97	\pm	0.17	$	&	$	3.07	\pm	0.20	$	&	$	5.0	\pm	0.8	$	&	B5IV	&	B7IV	&		&		&		\\
HD	16582	&	$	12	\pm	9	$	&	$	21500	\pm	500	$	&	$	4.10	\pm	0.10	$	&	$	7.26	\pm	0.07	$	&	$	3.60	\pm	0.10	$	&	$	7.9	\pm	0.3	$	&	B2IV	&	B2/B3V	&		&		&		\\
HD	17891	&	$	70	\pm	8	$	&	$	15150	\pm	400	$	&	$	3.55	\pm	0.10	$	&	$	7.97	\pm	0.05	$	&	$	3.11	\pm	0.10	$	&	$	5.0	\pm	0.3	$	&	B9	&	B6V	&		&		&		\\
HD	20340	&	$	200	\pm	18	$	&	$	18950	\pm	450	$	&	$	3.65	\pm	0.10	$	&	$	7.49	\pm	0.05	$	&	$	3.75	\pm	0.10	$	&	$	8.1	\pm	0.7	$	&	B3V	&	B3Ve	&	900	&	0.2	&	0.68	\\
HD	28248	&	$	30	\pm	11	$	&	$	20000	\pm	500	$	&	$	4.03	\pm	0.10	$	&	$	7.34	\pm	0.14	$	&	$	3.39	\pm	0.08	$	&	$	7.00	\pm	0.20	$	&	A1/A2V	&	B3V	&		&		&		\\
HD	29557	&	$	355	\pm	22	$	&	$	19000	\pm	450	$	&	$	4.00	\pm	0.10	$	&	$	7.42	\pm	0.09	$	&	$	3.31	\pm	0.09	$	&	$	6.70	\pm	0.10	$	&	B5Ib/Iib	&	B3Ve	&	2000	&	0.3	&	0.38	\\
HD	33453	&	$	290	\pm	22	$	&	$	13700	\pm	400	$	&	$	3.60	\pm	0.15	$	&	$	7.95	\pm	0.02	$	&	$	3.04	\pm	0.10	$	&	$	5.00	\pm	0.20	$	&	B8Vn	&	B8Vne	&	1500	&	0.3	&	0.79	\\
HD	33599	&	$	200	\pm	21	$	&	$	23200	\pm	500	$	&	$	4.03	\pm	0.10	$	&	$	7.12	\pm	0.13	$	&	$	3.78	\pm	0.10	$	&	$	9.2	\pm	0.3	$	&	B5psh	&	B2Vpe	&	700	&	0.1	&	0.35	\\
HD	35165	&	$	350	\pm	23	$	&	$	21500	\pm	500	$	&	$	3.77	\pm	0.10	$	&	$	7.34	\pm	0.10	$	&	$	3.90	\pm	0.10	$	&	$	8.9	\pm	0.3	$	&	B5IVnpe	&	B2Vnpe	&	1900	&	0.3	&	0.6	\\
HD	35468	&	$	40	\pm	12	$	&	$	21700	\pm	500	$	&	$	4.00	\pm	0.10	$	&	$	7.21	\pm	0.11	$	&	$	3.65	\pm	0.05	$	&	$	8.40	\pm	0.10	$	&	B2III	&	B2V	&		&		&		\\
HD	36012	&	$	180	\pm	13	$	&	$	21700	\pm	500	$	&	$	3.52	\pm	0.10	$	&	$	7.34	\pm	0.02	$	&	$	4.21	\pm	0.20	$	&	$	10	\pm	1	$	&	B5Vne	&	B2Vne	&	700	&	0.2	&	0.99	\\
HD	36861	&	$	55	\pm	10	$	&	$	34000	\pm	500	$	&	$	3.82	\pm	0.10	$	&	$	6.59	\pm	0.15	$	&	$	5.06	\pm	0.20	$	&	$	25	\pm	2	$	&	O8III	&	O8V	&		&		&		\\
HD	37023	&	$	50	\pm	10	$	&	$	24200	\pm	450	$	&	$	3.81	\pm	0.15	$	&	$	7.13	\pm	0.03	$	&	$	4.17	\pm	0.20	$	&	$	11	\pm	1	$	&	B0.5Vp	&	B1.5Vp	&		&		&		\\
HD	37490	&	$	170	\pm	15	$	&	$	19000	\pm	500	$	&	$	3.59	\pm	0.10	$	&	$	7.64	\pm	0.09	$	&	$	3.68	\pm	0.10	$	&	$	7.0	\pm	0.5	$	&	B3IIIe	&	B3Ve	&	700	&	0.2	&	0.94	\\
HD	37795	&	$	180	\pm	15	$	&	$	14200	\pm	400	$	&	$	3.50	\pm	0.10	$	&	$	7.97	\pm	0.12	$	&	$	3.07	\pm	0.20	$	&	$	5.0	\pm	0.5	$	&	B7IVe	&	B9Ve	&	1500	&	0.2	&	0.91	\\
HD	43122	&	$	265	\pm	18	$	&	$	25300	\pm	600	$	&	$	3.63	\pm	0.10	$	&	$	7.10	\pm	0.09	$	&	$	4.4	\pm	0.3	$	&	$	13	\pm	2	$	&	B8	&	B1Ve	&	1200	&	0.05	&	0.67	\\
HD	43285	&	$	237	\pm	11	$	&	$	16600	\pm	600	$	&	$	4.00	\pm	0.10	$	&	$	7.74	\pm	0.08	$	&	$	3.02	\pm	0.10	$	&	$	5.38	\pm	0.20	$	&	B6Ve	&	B6Ve	&	900	&	0.2	&	0.43	\\
HD	43544	&	$	260	\pm	22	$	&	$	21500	\pm	500	$	&	$	3.91	\pm	0.10	$	&	$	7.29	\pm	0.09	$	&	$	3.72	\pm	0.10	$	&	$	8.51	\pm	0.10	$	&	B2/B3V	&	B2/B3Ve	&	1200	&	0.2	&	0.45	\\
HD	43789	&	$	255	\pm	19	$	&	$	15100	\pm	400	$	&	$	3.70	\pm	0.15	$	&	$	7.86	\pm	0.05	$	&	$	3.16	\pm	0.10	$	&	$	5.5	\pm	0.3	$	&	B6V	&	B6/B7Ve	&	900	&	0.2	&	0.71	\\
HD	44743	&	$	20	\pm	7	$	&	$	24000	\pm	500	$	&	$	3.43	\pm	0.10	$	&	$	7.06	\pm	0.04	$	&	$	4.70	\pm	0.10	$	&	$	14.7	\pm	0.3	$	&	B1II/III	&	B1.5IV	&		&		&		\\
HD	44996	&	$	38	\pm	11	$	&	$	15200	\pm	500	$	&	$	3.90	\pm	0.10	$	&	$	7.85	\pm	0.07	$	&	$	2.96	\pm	0.15	$	&	$	5.00	\pm	0.20	$	&	B4V	&	B6/B7V	&		&		&		\\
HD	45871	&	$	275	\pm	15	$	&	$	20000	\pm	500	$	&	$	3.72	\pm	0.10	$	&	$	7.42	\pm	0.08	$	&	$	3.79	\pm	0.10	$	&	$	8.39	\pm	0.20	$	&	B4Vnpe	&	B3Ve	&	1500	&	0.3	&	0.59	\\
HD	46131	&	$	275	\pm	18	$	&	$	19500	\pm	600	$	&	$	4.05	\pm	0.15	$	&	$	7.34	\pm	0.06	$	&	$	3.39	\pm	0.10	$	&	$	7.00	\pm	0.20	$	&	B4V	&	B3Ve	&	1500	&	0.3	&	0.38	\\
HD	46380	&	$	293	\pm	30	$	&	$	21100	\pm	650	$	&	$	3.50	\pm	0.10	$	&	$	7.35	\pm	0.05	$	&	$	4.18	\pm	0.10	$	&	$	10.1	\pm	0.9	$	&	B2Vne	&	B2/B3IVne	&	1600	&	0.3	&	0.84	\\
HD	47839	&	$	70	\pm	10	$	&	$	25200	\pm	500	$	&	$	3.50	\pm	0.10	$	&	$	7.06	\pm	0.06	$	&	$	4.74	\pm	0.10	$	&	$	14.7	\pm	0.3	$	&	O7Ve	&	B1Ve	&	400	&	0.1	&	0.92	\\
HD	48099	&	$	115	\pm	12	$	&	$	30700	\pm	700	$	&	$	3.53	\pm	0.10	$	&	$	6.75	\pm	0.04	$	&	$	5.17	\pm	0.10	$	&	$	24	\pm	1	$	&	O6e	&	B0Ve	&	100	&	0.05	&	0.68	\\
HD	48282	&	$	188	\pm	18	$	&	$	18520	\pm	550	$	&	$	4.10	\pm	0.15	$	&	$	7.35	\pm	0.20	$	&	$	3.14	\pm	0.10	$	&	$	6.2	\pm	0.7	$	&	B3III	&	B3Ve	&	1200	&	0.2	&	0.32	\\
HD	49131	&	$	135	\pm	15	$	&	$	20000	\pm	500	$	&	$	3.60	\pm	0.10	$	&	$	7.42	\pm	0.15	$	&	$	4.06	\pm	0.20	$	&	$	9	\pm	1	$	&	B2III	&	B3V	&		&		&		\\
HD	49319	&	$	245	\pm	18	$	&	$	21000	\pm	500	$	&	$	3.40	\pm	0.15	$	&	$	7.20	\pm	0.05	$	&	$	4.39	\pm	0.20	$	&	$	12	\pm	1	$	&	B6Vnne	&	B2/B3IVne	&	1400	&	0.3	&	0.93	\\
HD	49330	&	$	200	\pm	10	$	&	$	27200	\pm	600	$	&	$	4.00	\pm	0.10	$	&	$	6.91	\pm	0.20	$	&	$	4.16	\pm	0.10	$	&	$	11.9	\pm	0.4	$	&	B0:nnpe	&	B1Vnpe	&	300	&	0.05	&	0.36	\\
HD	49336	&	$	220	\pm	15	$	&	$	16400	\pm	500	$	&	$	3.59	\pm	0.10	$	&	$	7.61	\pm	0.20	$	&	$	3.58	\pm	0.10	$	&	$	7.00	\pm	0.10	$	&	B4Vne	&	B6Vne	&	1100	&	0.2	&	0.79	\\
HD	50013	&	$	210	\pm	20	$	&	$	24100	\pm	500	$	&	$	3.58	\pm	0.10	$	&	$	7.05	\pm	0.02	$	&	$	4.53	\pm	0.05	$	&	$	14	\pm	1	$	&	B1.5IVe	&	B1.5Ve	&	600	&	0.1	&	0.67	\\
HD	50209	&	$	173	\pm	15	$	&	$	12500	\pm	450	$	&	$	4.00	\pm	0.15	$	&	$	8.03	\pm	0.20	$	&	$	2.55	\pm	0.30	$	&	$	4	\pm	0.9	$	&	B9Ve	&	B9Ve	&	200	&	0.05	&	0.45	\\
HD	50696	&	$	281	\pm	18	$	&	$	21700	\pm	600	$	&	$	3.50	\pm	0.15	$	&	$	7.20	\pm	0.10	$	&	$	4.43	\pm	0.25	$	&	$	12	\pm	1	$	&	B1:V:nne	&	B2Vnne	&	1400	&	0.3	&	1	\\
HD	50737	&	$	230	\pm	20	$	&	$	22200	\pm	600	$	&	$	4.00	\pm	0.15	$	&	$	7.17	\pm	0.10	$	&	$	3.71	\pm	0.10	$	&	$	8.7	\pm	0.3	$	&	B2Vnne	&	B2Vnne	&	800	&	0.2	&	0.33	\\
HD	50850	&	$	310	\pm	26	$	&	$	18900	\pm	600	$	&	$	4.20	\pm	0.10	$	&	$	7.02	\pm	0.20	$	&	$	3.08	\pm	0.09	$	&	$	6.18	\pm	0.10	$	&	B3:Vnne	&	B3:Vnne	&	1600	&	0.3	&	0.21	\\
HD	51309	&	$	28	\pm	9	$	&	$	18900	\pm	500	$	&	$	3.53	\pm	0.15	$	&	$	7.41	\pm	0.08	$	&	$	3.98	\pm	0.20	$	&	$	9	\pm	1	$	&	B3Ib/II	&	B3V	&		&		&		\\
HD	52159	&	$	110	\pm	17	$	&	$	17700	\pm	550	$	&	$	3.50	\pm	0.15	$	&	$	7.56	\pm	0.08	$	&	$	3.76	\pm	0.20	$	&	$	7.7	\pm	0.8	$	&	B5Vne	&	B3Vne	&	700	&	0.1	&	0.91	\\
HD	52244	&	$	210	\pm	22	$	&	$	22170	\pm	550	$	&	$	3.50	\pm	0.10	$	&	$	7.20	\pm	0.04	$	&	$	4.43	\pm	0.15	$	&	$	12	\pm	1	$	&	B2:III:npe	&	B2IVnpe	&	800	&	0.2	&	0.99	\\
HD	55606	&	$	335	\pm	20	$	&	$	28700	\pm	550	$	&	$	4.10	\pm	0.10	$	&	$	6.68	\pm	0.25	$	&	$	4.24	\pm	0.12	$	&	$	14	\pm	1	$	&	B1:V:nnpe	&	B0.5Vnnpe	&	100	&	0.05	&	0.3	\\
HD	58715	&	$	230	\pm	20	$	&	$	13100	\pm	400	$	&	$	3.61	\pm	0.10	$	&	$	7.95	\pm	0.07	$	&	$	3.04	\pm	0.10	$	&	$	5.00	\pm	0.10	$	&	B8Ve	&	B8Ve	&	700	&	0.1	&	0.79	\\
HD	59868	&	$	200	\pm	18	$	&	$	17000	\pm	500	$	&	$	3.82	\pm	0.10	$	&	$	7.70	\pm	0.06	$	&	$	3.25	\pm	0.30	$	&	$	6.1	\pm	1	$	&	B8IV/V	&	B3V	&		&		&		\\
HD	63150	&	$	260	\pm	15	$	&	$	25150	\pm	450	$	&	$	3.75	\pm	0.15	$	&	$	7.12	\pm	0.10	$	&	$	4.30	\pm	0.20	$	&	$	12	\pm	1	$	&	B0.5Vnne	&	B1Ve	&	1100	&	0.05	&	0.62	\\
HD	67698	&	$	90	\pm	11	$	&	$	17400	\pm	500	$	&	$	3.55	\pm	0.10	$	&	$	7.63	\pm	0.14	$	&	$	3.65	\pm	0.10	$	&	$	7	\pm	1	$	&	B3III/IV	&	B3Ve	&	700	&	0.1	&	1	\\
HD	70461	&	$	266	\pm	15	$	&	$	15940	\pm	550	$	&	$	3.95	\pm	0.10	$	&	$	7.79	\pm	0.10	$	&	$	2.91	\pm	0.10	$	&	$	5.0	\pm	0.5	$	&	B5	&	B6Ve	&	900	&	0.2	&	0.48	\\
HD	74280	&	$	120	\pm	15	$	&	$	18000	\pm	500	$	&	$	4.00	\pm	0.10	$	&	$	7.52	\pm	0.12	$	&	$	3.18	\pm	0.08	$	&	$	6.2	\pm	0.3	$	&	B3V	&	B3V	&		&		&		\\
HD	79447	&	$	13	\pm	8	$	&	$	18900	\pm	500	$	&	$	3.50	\pm	0.10	$	&	$	7.43	\pm	0.03	$	&	$	3.94	\pm	0.08	$	&	$	8.78	\pm	0.3	$	&	B3III	&	B3V	&		&		&		\\
HD	90177	&	$	10	\pm	5	$	&	$	14100	\pm	400	$	&	$	3.58	\pm	0.10	$	&	$	7.91	\pm	0.03	$	&	$	3.1	\pm	0.3	$	&	$	5.29	\pm	1	$	&	B2evar	&	B7Ve	&	300	&	0.05	&	0.78	\\
HD	97991	&	$150	\pm	10$	&	$21000	\pm	400$	&	$4.01	\pm	0.10$	&	$7.45	\pm	0.10$	&	$3.56	\pm	0.10$	&	$7.80	\pm	0.10$	&	B2/B3V	&	B2/B3V	&	&	&	\\
HD	100546	&	$	85	\pm	20	$	&	$	11750	\pm	600	$	&	$	3.50	\pm	0.10	$	&	$	8.21	\pm	0.10	$	&	$	2.69	\pm	0.25	$	&	$	4	\pm	1	$	&	B9Vne	&	B9Vne	&	100	&	0.05	&	0.93	\\
HD	100889	&	$	180	\pm	18	$	&	$	12100	\pm	400	$	&	$	4.00	\pm	0.10	$	&	$	8.24	\pm	0.14	$	&	$	2.26	\pm	0.10	$	&	$	3.41	\pm	0.12	$	&	B9.5Vn	&	B9.5V	&		&		&		\\
HD	104582	&	$	159	\pm	16	$	&	$	13500	\pm	500	$	&	$	4.00	\pm	0.15	$	&	$	8.03	\pm	0.10	$	&	$	2.55	\pm	0.17	$	&	$	4.0	\pm	0.8	$	&	B8/B9II/III	&	B8/B9Ve	&	200	&	0.05	&	0.45	\\
HD	105435	&	$	240	\pm	32	$	&	$	22230	\pm	650	$	&	$	4.00	\pm	0.10	$	&	$	7.17	\pm	0.15	$	&	$	3.71	\pm	0.10	$	&	$	8.72	\pm	0.3	$	&	B2IVne	&	B2Vne	&	900	&	0.2	&	0.34	\\
HD	105521	&	$	160	\pm	18	$	&	$	19200	\pm	550	$	&	$	3.50	\pm	0.10	$	&	$	7.41	\pm	0.10	$	&	$	3.98	\pm	0.25	$	&	$	8.98	\pm	1	$	&	B3IVe	&	B3IVe	&	500	&	0.2	&	0.85	\\
HD	105937	&	$	110	\pm	20	$	&	$	19500	\pm	600	$	&	$	3.95	\pm	0.10	$	&	$	7.44	\pm	0.04	$	&	$	3.45	\pm	0.07	$	&	$	7.03	\pm	0.17	$	&	B3V	&	B3V	&		&		&		\\
HD	106309	&	$	250	\pm	28	$	&	$	22000	\pm	500	$	&	$	3.58	\pm	0.10	$	&	$	7.42	\pm	0.10	$	&	$	4.06	\pm	0.07	$	&	$	9	\pm	1	$	&	B2IV:ne	&	B2IVne	&	1200	&	0.2	&	0.9	\\
HD	106793	&	$	277	\pm	20	$	&	$	13880	\pm	450	$	&	$	3.53	\pm	0.15	$	&	$	7.96	\pm	0.10	$	&	$	3.06	\pm	0.10	$	&	$	5.0	\pm	0.5	$	&	B8/B9II/IIIe	&	B8/B9IVe	&	1400	&	0.3	&	0.85	\\
HD	110432	&	$	400	\pm	30	$	&	$	27100	\pm	650	$	&	$	3.00	\pm	0.10	$	&	$	6.66	\pm	0.04	$	&	$	5.63	\pm	0.21	$	&	$	34.26	\pm	5	$	&	B2pe	&	B0.5IVpe	&	1600	&	0.05	&	0.91	\\
HD	110699	&	$	150	\pm	15	$	&	$	15000	\pm	500	$	&	$	3.62	\pm	0.10	$	&	$	7.98	\pm	0.02	$	&	$	3.12	\pm	0.20	$	&	$	5.0	\pm	0.4	$	&	B9.5:V:n	&	B6Vne	&	400	&	0.05	&	0.92	\\
HD	112078	&	$	300	\pm	20	$	&	$	17300	\pm	450	$	&	$	3.62	\pm	0.10	$	&	$	7.64	\pm	0.03	$	&	$	3.68	\pm	0.03	$	&	$	7.00	\pm	0.13	$	&	B4Vne	&	B3Vne	&	1800	&	0.36	&	0.94	\\
HD	112091	&	$	230	\pm	20	$	&	$	16500	\pm	550	$	&	$	3.50	\pm	0.10	$	&	$	7.63	\pm	0.10	$	&	$	3.62	\pm	0.16	$	&	$	7.0	\pm	0.7	$	&	B5Vne	&	B5Vne	&	1200	&	0.25	&	0.95	\\
HD	112107	&	$	200	\pm	19	$	&	$	12000	\pm	400	$	&	$	3.58	\pm	0.10	$	&	$	8.21	\pm	0.15	$	&	$	2.74	\pm	0.14	$	&	$	4.04	\pm	0.18	$	&	B9.5Vn	&	B9.5Vne	&	400	&	0.05	&	0.99	\\
HD	112512	&	$	199	\pm	20	$	&	$	14180	\pm	400	$	&	$	3.69	\pm	0.15	$	&	$	7.93	\pm	0.06	$	&	$	3.02	\pm	0.15	$	&	$	5.0	\pm	0.5	$	&	B7III	&	B7Ve	&	600	&	0.1	&	0.67	\\
HD	113120	&	$	320	\pm	20	$	&	$	22800	\pm	500	$	&	$	3.42	\pm	0.10	$	&	$	7.11	\pm	0.05	$	&	$	4.55	\pm	0.06	$	&	$	14	\pm	1	$	&	B1.5IIIne	&	B2IVne	&	1700	&	0.2	&	0.87	\\
HD	118094	&	$	239	\pm	30	$	&	$	13500	\pm	550	$	&	$	3.90	\pm	0.15	$	&	$	8.06	\pm	0.04	$	&	$	2.65	\pm	0.08	$	&	$	4.18	\pm	0.20	$	&	B8Vn...	&	B8Vne	&	600	&	0.1	&	0.57	\\
HD	119423	&	$	180	\pm	15	$	&	$	17000	\pm	600	$	&	$	3.95	\pm	0.15	$	&	$	7.79	\pm	0.13	$	&	$	2.91	\pm	0.20	$	&	$	5.0	\pm	0.4	$	&	B4:Vne	&	B3Vne	&	700	&	0.2	&	0.43	\\
HD	119682	&	$	200	\pm	20	$	&	$	31910	\pm	550	$	&	$	4.00	\pm	0.10	$	&	$	6.59	\pm	0.10	$	&	$	4.78	\pm	0.10	$	&	$	20	\pm	1	$	&	O+...	&	B0Ve	&	100	&	0.05	&	0.37	\\
HD	120324	&	$	125	\pm	16	$	&	$	22500	\pm	550	$	&	$	3.75	\pm	0.15	$	&	$	7.28	\pm	0.25	$	&	$	4.03	\pm	0.11	$	&	$	9.95	\pm	0.2	$	&	B2Vnpe	&	B2Vnpe	&	700	&	0.1	&	0.6	\\
HD	120991	&	$	130	\pm	16	$	&	$	22000	\pm	500	$	&	$	3.57	\pm	0.10	$	&	$	7.42	\pm	0.02	$	&	$	4.06	\pm	0.13	$	&	$	8.9	\pm	0.5	$	&	B2IIIe	&	B2Ve	&	500	&	0.1	&	0.92	\\
\hline
\end{tabular}}
\end{sideways}
\end{centering}
\end{table*}

\eject

\addtocounter{table}{-1}%

\begin{table*}
\begin{centering}
\rotcaption{Continued.}
\begin{sideways}
\tiny{
\begin{tabular}{l|ccccccllccc}
\hline\hline
Object  &  \vsini (\kps)   &  \teff (K)  &  \logg (dex)  & \lage (yr)  &  \logl &  \mmsun &  Sp.T.(Simbad) &  Sp.T.(this work) & $\Delta$\teff (K) & $\Delta$\logg (dex) & $\tau/\tau_{\rm M.S.}$ \\
\hline\hline
HD	124639	&	$	237	\pm	30	$	&	$	12700	\pm	600	$	&	$	3.50	\pm	0.10	$	&	$	8.05	\pm	0.09	$	&	$	2.96	\pm	0.14	$	&	$	4.7	\pm	0.3	$	&	B8Ve	&	B9IVe	&	900	&	0.2	&	0.91	\\
HD	125924	&	$75	\pm	12$	&	$21000	\pm	400$	&	$4.10	\pm	0.15$	&	$7.10	\pm	0.15$	&	$3.77	\pm	0.16$	&	$8.90	\pm	0.10$	&	B2IV	&	B2IV	&	&	&	\\
HD	126527	&	$	235	\pm	25	$	&	$	13200	\pm	450	$	&	$	3.67	\pm	0.10	$	&	$	8.05	\pm	0.08	$	&	$	2.85	\pm	0.17	$	&	$	4.5	\pm	0.3	$	&	B8V	&	B9Ve	&	600	&	0.1	&	0.7	\\
HD	126986	&	$	70	\pm	15	$	&	$	12000	\pm	400	$	&	$	3.54	\pm	0.10	$	&	$	8.14	\pm	0.11	$	&	$	2.80	\pm	0.17	$	&	$	4.3	\pm	0.4	$	&	B9.5Vnn...	&	B9IVne	&	200	&	0.05	&	0.86	\\
HD	127112	&	$	115	\pm	20	$	&	$	14000	\pm	500	$	&	$	3.63	\pm	0.15	$	&	$	8.06	\pm	0.20	$	&	$	3.00	\pm	0.20	$	&	$	4.7	\pm	0.3	$	&	B7III	&	B8Ve	&	300	&	0.05	&	0.88	\\
HD	127208	&	$	150	\pm	20	$	&	$	12000	\pm	500	$	&	$	3.55	\pm	0.10	$	&	$	8.22	\pm	0.15	$	&	$	2.73	\pm	0.10	$	&	$	4.00	\pm	0.18	$	&	B8Ve	&	B9IVe	&	200	&	0.05	&	0.99	\\
HD	127381	&	$	80	\pm	14	$	&	$	23000	\pm	550	$	&	$	4.02	\pm	0.10	$	&	$	7.13	\pm	0.13	$	&	$	3.76	\pm	0.06	$	&	$	9	\pm	0.5	$	&	B2III	&	B1/B2V	&		&		&		\\
HD	127972	&	$	310	\pm	20	$	&	$	20500	\pm	600	$	&	$	3.80	\pm	0.10	$	&	$	7.36	\pm	0.1	$	&	$	3.76	\pm	0.16	$	&	$	8.5	\pm	0.9	$	&	B1.5Vne	&	B2Ve	&	1500	&	0.3	&	0.53	\\
HD	129956	&	$	80	\pm	15	$	&	$	11500	\pm	400	$	&	$	3.61	\pm	0.10	$	&	$	8.19	\pm	0.03	$	&	$	2.66	\pm	0.06	$	&	$	4	\pm	0.13	$	&	B9.5V	&	B9.5V	&		&		&		\\
HD	130437	&	$	335	\pm	25	$	&	$	24700	\pm	500	$	&	$	3.52	\pm	0.15	$	&	$	7.15	\pm	0.02	$	&	$	4.59	\pm	0.09	$	&	$	13.2	\pm	0.7	$	&	O+...	&	B1Ve	&	1700	&	0.2	&	0.93	\\
HD	130534	&	$	100	\pm	20	$	&	$	14700	\pm	400	$	&	$	3.55	\pm	0.15	$	&	$	7.97	\pm	0.13	$	&	$	3.11	\pm	0.20	$	&	$	5	\pm	1	$	&	B3III	&	B7Ve	&	200	&	0.05	&	1	\\
HD	131168	&	$	185	\pm	22	$	&	$	22000	\pm	500	$	&	$	3.70	\pm	0.10	$	&	$	7.34	\pm	0.09	$	&	$	3.90	\pm	0.11	$	&	$	8.9	\pm	0.4	$	&	B3Ve	&	B2Ve	&	800	&	0.1	&	0.6	\\
HD	132947	&	$	150	\pm	20	$	&	$	13000	\pm	500	$	&	$	4.05	\pm	0.10	$	&	$	8.06	\pm	0.08	$	&	$	2.35	\pm	0.23	$	&	$	3.7	\pm	0.4	$	&	A0	&	B9V	&		&		&		\\
HD	134401	&	$	375	\pm	25	$	&	$	22200	\pm	500	$	&	$	3.77	\pm	0.10	$	&	$	7.32	\pm	0.12	$	&	$	3.95	\pm	0.11	$	&	$	9.4	\pm	0.4	$	&	B2Vne	&	B2Vne	&	2000	&	0.35	&	0.59	\\
HD	134481	&	$	146	\pm	19	$	&	$	11700	\pm	500	$	&	$	4.00	\pm	0.10	$	&	$	8.29	\pm	0.08	$	&	$	2.18	\pm	0.20	$	&	$	3.25	\pm	0.10	$	&	B9.5Vne	&	B9.5Vne	&	100	&	0.05	&	0.44	\\
HD	134671	&	$	250	\pm	19	$	&	$	14700	\pm	400	$	&	$	3.81	\pm	0.15	$	&	$	7.90	\pm	0.05	$	&	$	2.99	\pm	0.09	$	&	$	5	\pm	0.2	$	&	B7II	&	B7Ve	&	800	&	0.2	&	0.62	\\
HD	135734	&	$	280	\pm	20	$	&	$	13470	\pm	500	$	&	$	3.80	\pm	0.10	$	&	$	8.05	\pm	0.05	$	&	$	2.78	\pm	0.10	$	&	$	4.42	\pm	0.20	$	&	B8Ve	&	B8Ve	&	900	&	0.2	&	0.59	\\
HD	136968	&	$	207	\pm	19	$	&	$	16350	\pm	400	$	&	$	3.57	\pm	0.10	$	&	$	7.7	\pm	0.15	$	&	$	3.48	\pm	0.10	$	&	$	7	\pm	1	$	&	B5Vne	&	B5Vne	&	1000	&	0.2	&	0.79	\\
HD	137387	&	$	250	\pm	21	$	&	$	21500	\pm	500	$	&	$	3.90	\pm	0.10	$	&	$	7.29	\pm	0.18	$	&	$	3.73	\pm	0.15	$	&	$	8.6	\pm	0.9	$	&	B1npe	&	B2Vnpe	&	1100	&	0.2	&	0.43	\\
HD	137518	&	$	300	\pm	25	$	&	$	25900	\pm	400	$	&	$	3.82	\pm	0.15	$	&	$	7.07	\pm	0.17	$	&	$	4.25	\pm	0.24	$	&	$	12	\pm	1	$	&	B1/B2IIIn...	&	B1Vne	&	1000	&	0.1	&	0.54	\\
HD	142237	&	$	345	\pm	27	$	&	$	26000	\pm	500	$	&	$	3.65	\pm	0.10	$	&	$	6.98	\pm	0.14	$	&	$	4.58	\pm	0.10	$	&	$	14.8	\pm	0.4	$	&	B2Vne	&	B1Vne	&	300	&	0.1	&	0.59	\\
HD	142349	&	$	250	\pm	25	$	&	$	16600	\pm	550	$	&	$	3.70	\pm	0.10	$	&	$	7.58	\pm	0.08	$	&	$	3.56	\pm	0.08	$	&	$	7.00	\pm	0.20	$	&	B5IV	&	B5Ve	&	700	&	0.2	&	0.7	\\
HD	143545	&	$	370	\pm	30	$	&	$	23000	\pm	550	$	&	$	3.70	\pm	0.15	$	&	$	7.19	\pm	0.23	$	&	$	4.19	\pm	0.20	$	&	$	11.18	\pm	0.20	$	&	B1/B2ne	&	B1/B2Vne	&	2100	&	0.3	&	0.6	\\
HD	143578	&	$	164	\pm	16	$	&	$	13500	\pm	400	$	&	$	3.55	\pm	0.10	$	&	$	7.96	\pm	0.05	$	&	$	3.07	\pm	0.10	$	&	$	5.0	\pm	0.5	$	&	B8IV/V	&	B8IV/Ve	&	400	&	0.1	&	0.86	\\
HD	143700	&	$	230	\pm	20	$	&	$	23000	\pm	500	$	&	$	3.60	\pm	0.10	$	&	$	7.15	\pm	0.36	$	&	$	4.33	\pm	0.20	$	&	$	12	\pm	1	$	&	B2/B3III:ne	&	B1/B2Vne	&	1000	&	0.1	&	0.67	\\
HD	144555	&	$	275	\pm	23	$	&	$	24000	\pm	450	$	&	$	3.73	\pm	0.10	$	&	$	7.12	\pm	0.18	$	&	$	4.30	\pm	0.08	$	&	$	12	\pm	1	$	&	B2Ib:ne	&	B1Vne	&	1100	&	0.2	&	0.62	\\
HD	146531	&	$	260	\pm	25	$	&	$	17000	\pm	500	$	&	$	3.67	\pm	0.10	$	&	$	7.58	\pm	0.21	$	&	$	3.56	\pm	0.10	$	&	$	7.00	\pm	0.16	$	&	B6III	&	B3Ve	&	1300	&	0.3	&	0.69	\\
HD	149757	&	$	340	\pm	20	$	&	$	27000	\pm	650	$	&	$	4.00	\pm	0.10	$	&	$	6.91	\pm	0.20	$	&	$	4.16	\pm	0.10	$	&	$	12	\pm	1	$	&	O9V	&	B0.5Ve	&	200	&	0.1	&	0.37	\\
HD	150193	&	$	65	\pm	15	$	&	$	11000	\pm	400	$	&	$	3.58	\pm	0.10	$	&	$	8.37	\pm	0.15	$	&	$	2.54	\pm	0.08	$	&	$	3.62	\pm	0.10	$	&	A1Ve	&	B9.5Ve	&	300	&	0.05	&	0.99	\\
HD	150288	&	$	250	\pm	23	$	&	$	22000	\pm	450	$	&	$	3.58	\pm	0.10	$	&	$	7.42	\pm	0.09	$	&	$	4.06	\pm	0.12	$	&	$	8.9	\pm	0.5	$	&	B3V	&	B2Ve	&	1200	&	0.2	&	0.95	\\
HD	150422	&	$	270	\pm	26	$	&	$	25100	\pm	400	$	&	$	3.83	\pm	0.10	$	&	$	7.07	\pm	0.03	$	&	$	4.25	\pm	0.08	$	&	$	11.9	\pm	0.3	$	&	B1/B2ne	&	B1Vne	&	1100	&	0.1	&	0.52	\\
HD	151113	&	$	180	\pm	20	$	&	$	21700	\pm	600	$	&	$	3.71	\pm	0.10	$	&	$	7.34	\pm	0.11	$	&	$	3.90	\pm	0.10	$	&	$	8.9	\pm	0.5	$	&	B5II/III	&	B2Ve	&	800	&	0.2	&	0.61	\\
HD	152060	&	$	105	\pm	12	$	&	$	22000	\pm	500	$	&	$	3.64	\pm	0.15	$	&	$	7.15	\pm	0.18	$	&	$	4.33	\pm	0.15	$	&	$	11.9	\pm	0.4	$	&	B1III	&	B2Ve	&	700	&	0.1	&	0.67	\\
HD	152478	&	$	295	\pm	22	$	&	$	19800	\pm	500	$	&	$	3.75	\pm	0.10	$	&	$	7.34	\pm	0.09	$	&	$	3.90	\pm	0.08	$	&	$	8.99	\pm	0.2	$	&	B3Vnpe	&	B3Vnpe	&	1600	&	0.3	&	0.61	\\
HD	152979	&	$	190	\pm	20	$	&	$	21000	\pm	500	$	&	$	3.51	\pm	0.10	$	&	$	7.35	\pm	0.08	$	&	$	4.18	\pm	0.08	$	&	$	10.04	\pm	0.10	$	&	B2IV	&	B2IVe	&	900	&	0.2	&	0.87	\\
HD	153199	&	$	190	\pm	19	$	&	$	20200	\pm	500	$	&	$	3.68	\pm	0.10	$	&	$	7.37	\pm	0.09	$	&	$	3.93	\pm	0.10	$	&	$	8.9	\pm	0.5	$	&	B3II/III	&	B2/B3Ve	&	900	&	0.2	&	0.69	\\
HD	154154	&	$	290	\pm	25	$	&	$	25000	\pm	500	$	&	$	3.65	\pm	0.15	$	&	$	7.15	\pm	0.03	$	&	$	4.34	\pm	0.10	$	&	$	12.01	\pm	0.20	$	&	B2Vnne	&	B1Vnne	&	1400	&	0.05	&	0.7	\\
HD	155851	&	$	335	\pm	30	$	&	$	30000	\pm	700	$	&	$	3.65	\pm	0.10	$	&	$	6.81	\pm	0.14	$	&	$	4.93	\pm	0.15	$	&	$	19.89	\pm	0.20	$	&	B0Vn	&	B0Vne	&	1000	&	0.1	&	0.59	\\
HD	156325	&	$	180	\pm	24	$	&	$	15700	\pm	500	$	&	$	3.57	\pm	0.15	$	&	$	7.84	\pm	0.08	$	&	$	3.37	\pm	0.20	$	&	$	6	\pm	1	$	&	B5Vne	&	B6Vne	&	700	&	0.1	&	0.99	\\
HD	156702	&	$	265	\pm	26	$	&	$	24700	\pm	500	$	&	$	4.01	\pm	0.10	$	&	$	7.03	\pm	0.12	$	&	$	3.97	\pm	0.10	$	&	$	10.57	\pm	0.20	$	&	B5	&	B1Ve	&	800	&	0.1	&	0.35	\\
HD	158427	&	$	270	\pm	25	$	&	$	22150	\pm	550	$	&	$	3.55	\pm	0.10	$	&	$	7.17	\pm	0.10	$	&	$	4.36	\pm	0.10	$	&	$	11.90	\pm	0.20	$	&	B2Vne	&	B2Vne	&	1400	&	0.2	&	0.74	\\
HD	159489	&	$	170	\pm	18	$	&	$	25700	\pm	500	$	&	$	4.04	\pm	0.10	$	&	$	6.91	\pm	0.14	$	&	$	4.16	\pm	0.10	$	&	$	11.97	\pm	0.20	$	&	B3IV	&	B1Ve	&	700	&	0.05	&	0.37	\\
HD	160202	&	$	220	\pm	21	$	&	$	16700	\pm	500	$	&	$	3.55	\pm	0.10	$	&	$	7.63	\pm	0.17	$	&	$	3.65	\pm	0.10	$	&	$	7.00	\pm	0.10	$	&	B7Ve	&	B5Ve	&	1200	&	0.2	&	0.99	\\
HD	161774	&	$	100	\pm	10	$	&	$	13200	\pm	500	$	&	$	3.65	\pm	0.15	$	&	$	8.22	\pm	0.10	$	&	$	2.78	\pm	0.10	$	&	$	4.00	\pm	0.20	$	&	B5V:nne	&	B8Vnne	&	200	&	0.05	&	0.93	\\
HD	164284	&	$	220	\pm	20	$	&	$	22200	\pm	400	$	&	$	3.60	\pm	0.15	$	&	$	7.39	\pm	0.16	$	&	$	4.13	\pm	0.08	$	&	$	9	\pm	1	$	&	B2Ve	&	B2Ve	&	1000	&	0.2	&	0.91	\\
HD	164816	&	$	95	\pm	15	$	&	$	20000	\pm	500	$	&	$	3.53	\pm	0.10	$	&	$	7.44	\pm	0.08	$	&	$	4.01	\pm	0.05	$	&	$	8.87	\pm	0.60	$	&	O+...	&	B3Ve	&	500	&	0.1	&	0.81	\\
HD	164906	&	$	255	\pm	20	$	&	$	30700	\pm	550	$	&	$	3.95	\pm	0.15	$	&	$	6.67	\pm	0.20	$	&	$	4.64	\pm	0.15	$	&	$	18	\pm	1	$	&	O+...	&	B0Ve	&	100	&	0.05	&	0.34	\\
HD	164947	&	$	120	\pm	20	$	&	$	23000	\pm	500	$	&	$	4	\pm	0.10	$	&	$	7.13	\pm	0.20	$	&	$	3.76	\pm	0.15	$	&	$	9.00	\pm	0.20	$	&	B2V:n...	&	B1/B2Vne	&	700	&	0.1	&	0.38	\\
HD	165052	&	$	125	\pm	22	$	&	$	37500	\pm	550	$	&	$	4.10	\pm	0.10	$	&	$	5.97	\pm	0.04	$	&	$	4.9	\pm	0.3	$	&	$	25	\pm	3	$	&	O6.5V	&	O6.5V	&		&		&		\\
HD	166566	&	$	50	\pm	10	$	&	$	22000	\pm	500	$	&	$	3.60	\pm	0.10	$	&	$	7.40	\pm	0.09	$	&	$	4.11	\pm	0.10	$	&	$	9.3	\pm	0.3	$	&	B1III:ne	&	B2Ve	&	400	&	0.1	&	0.9	\\
HD	170235	&	$	170	\pm	18	$	&	$	26200	\pm	500	$	&	$	3.50	\pm	0.10	$	&	$	7.06	\pm	0.12	$	&	$	4.74	\pm	0.10	$	&	$	15	\pm	1	$	&	B2Vnne	&	B1Vnne	&	100	&	0.05	&	0.92	\\
HD	170835	&	$	260	\pm	23	$	&	$	22000	\pm	500	$	&	$	3.69	\pm	0.10	$	&	$	7.24	\pm	0.09	$	&	$	4.11	\pm	0.10	$	&	$	10.6	\pm	0.3	$	&	B5IV	&	B2Ve	&	1400	&	0.2	&	0.62	\\
HD	171054	&	$	40	\pm	10	$	&	$	22000	\pm	550	$	&	$	4.00	\pm	0.15	$	&	$	7.18	\pm	0.09	$	&	$	3.68	\pm	0.10	$	&	$	8.6	\pm	0.3	$	&	B2II	&	B2Ve	&	600	&	0.1	&	0.34	\\
HD	171219	&	$	190	\pm	25	$	&	$	13500	\pm	500	$	&	$	3.80	\pm	0.15	$	&	$	8.05	\pm	0.11	$	&	$	2.78	\pm	0.20	$	&	$	4.4	\pm	0.3	$	&	B8	&	B8V	&		&		&		\\
HD	172256	&	$	310	\pm	20	$	&	$	24000	\pm	500	$	&	$	4.02	\pm	0.10	$	&	$	7.13	\pm	0.18	$	&	$	3.76	\pm	0.10	$	&	$	9.0	\pm	0.5	$	&	B5II	&	B1Ve	&	1200	&	0.2	&	0.3	\\
HD	173948	&	$	130	\pm	10	$	&	$	22000	\pm	500	$	&	$	3.65	\pm	0.10	$	&	$	7.37	\pm	0.05	$	&	$	3.93	\pm	0.10	$	&	$	8.9	\pm	0.5	$	&	B2II-IIIe	&	B2Ve	&	700	&	0.1	&	0.68	\\
HD	174705	&	$	210	\pm	20	$	&	$	22150	\pm	400	$	&	$	3.6	\pm	0.10	$	&	$	7.40	\pm	0.10	$	&	$	4.12	\pm	0.10	$	&	$	9.4	\pm	0.3	$	&	B2Vne	&	B2Vne	&	900	&	0.1	&	0.94	\\
HD	179253	&	$	120	\pm	13	$	&	$	14800	\pm	500	$	&	$	3.62	\pm	0.10	$	&	$	7.99	\pm	0.06	$	&	$	3.10	\pm	0.10	$	&	$	5.00	\pm	0.10	$	&	B5V	&	B7Ve	&	300	&	0.1	&	0.89	\\
HD	183914	&	$	220	\pm	16	$	&	$	13200	\pm	600	$	&	$	4	\pm	0.15	$	&	$	8.03	\pm	0.35	$	&	$	2.55	\pm	0.15	$	&	$	4.0	\pm	0.8	$	&	B8Ve	&	B8Ve	&	100	&	0.05	&	0.43	\\
HD	184279	&	$	200	\pm	22	$	&	$	30400	\pm	600	$	&	$	3.90	\pm	0.10	$	&	$	6.86	\pm	0.07	$	&	$	4.49	\pm	0.10	$	&	$	15	\pm	1	$	&	B0.5IV	&	B0V	&		&		&		\\
HD	185037	&	$	280	\pm	20	$	&	$	13000	\pm	500	$	&	$	4.00	\pm	0.10	$	&	$	8.03	\pm	0.05	$	&	$	2.55	\pm	0.15	$	&	$	4.0	\pm	0.6	$	&	B8Vne	&	B8Vne	&	700	&	0.2	&	0.42	\\
HD	198001	&	$	90	\pm	10	$	&	$	11200	\pm	400	$	&	$	4.01	\pm	0.10	$	&	$	8.34	\pm	0.14	$	&	$	2.07	\pm	0.10	$	&	$	3.03	\pm	0.10	$	&	A1.5V	&	B9.5V	&		&		&		\\
HD	205637	&	$	242	\pm	21	$	&	$	18850	\pm	400	$	&	$	3.95	\pm	0.10	$	&	$	7.44	\pm	0.05	$	&	$	3.44	\pm	0.10	$	&	$	7.0	\pm	0.5	$	&	B3V:p	&	B3V	&		&		&		\\
HD	208886	&	$	200	\pm	20	$	&	$	15000	\pm	450	$	&	$	4.01	\pm	0.10	$	&	$	7.79	\pm	0.09	$	&	$	2.91	\pm	0.20	$	&	$	5.0	\pm	0.5	$	&	B5III	&	B7Ve	&	400	&	0.1	&	0.49	\\
HD	217891	&	$	90	\pm	15	$	&	$	15500	\pm	600	$	&	$	4.00	\pm	0.10	$	&	$	7.79	\pm	0.09	$	&	$	2.91	\pm	0.16	$	&	$	5.0	\pm	0.5	$	&	B6Ve	&	B6Ve	&	200	&	0.1	&	0.44	\\
HD	224686	&	$	300	\pm	20	$	&	$	13000	\pm	500	$	&	$	3.90	\pm	0.10	$	&	$	8.09	\pm	0.09	$	&	$	2.59	\pm	0.10	$	&	$	4.00	\pm	0.20	$	&	B9IV	&	B8V	&		&		&		\\
HD	316341	&	$	120	\pm	15	$	&	$	30000	\pm	500	$	&	$	3.54	\pm	0.10	$	&	$	6.85	\pm	0.55	$	&	$	4.94	\pm	0.20	$	&	$	19	\pm	2	$	&	O+...	&	B0Ve	&	200	&	0.1	&	0.68	\\
HD	316587	&	$	190	\pm	15	$	&	$	25700	\pm	500	$	&	$	3.90	\pm	0.10	$	&	$	7	\pm	0.15	$	&	$	4.20	\pm	0.10	$	&	$	11.9	\pm	0.6	$	&	B1:V:ne...	&	B1Vne	&	400	&	0.05	&	0.43	\\
HD	322422	&	$	170	\pm	15	$	&	$	26700	\pm	550	$	&	$	3.65	\pm	0.15	$	&	$	6.98	\pm	0.21	$	&	$	4.58	\pm	0.15	$	&	$	14.8	\pm	0.3	$	&	B0.5IIIe	&	B1Ve	&	200	&	0.05	&	0.59	\\
HD	330950	&	$	60	\pm	10	$	&	$	26700	\pm	500	$	&	$	4.00	\pm	0.10	$	&	$	6.91	\pm	0.21	$	&	$	4.16	\pm	0.15	$	&	$	11.9	\pm	0.3	$	&	O+...	&	B1Ve	&	100	&	0.05	&	0.38	\\
\hline
\end{tabular}}
\end{sideways}
\end{centering}
\end{table*}

\eject

\bsp

\label{lastpage}

\end{document}